\documentclass[superscriptaddress,groupedaddress,nofootnoteinbib,12pt]{article}  % for review and submission

\usepackage{graphicx}  % needed for figures
\usepackage{dcolumn}   % needed for some tables
\usepackage{bm}        % for math
\usepackage{jcappub}

\def\be{\begin{equation}}
\def\ee{\end{equation}}
\def\ba{\begin{eqnarray}}
\def\ea{\end{eqnarray}}
\def\beq{\begin{eqnarray}}
\def\eeq{\end{eqnarray}}

\def\mpl{M_{\rm p}}
\def\mpf{M_{\rm f}}

\def\d{\mathrm{d}}
\def\p{{\cal P}}

\def\L*{{\cal L}_*}
\def\L{\mathcal{L}}
\def\({\left(}
\def\){\right)}

\def\nn{\nonumber}
\def\p{\partial}
\def\mn{_{\mu \nu}}
\def\stu{St\"uckelberg }
\def\p{\partial}

\def\<{\langle}
\def\>{\rangle}

\def\cs2{c_{s}^{2}}

 \def\al{\alpha}
 \def\b{\beta}
 
 \def\de{\delta}
 
 \def\ep{\varepsilon}

 \def\La{\Lambda}
 \def\si{\sigma}

 \def\om{\omega}

 \def\p{\partial}

 \def\be   {\begin{equation}}   \def\ee   {\end{equation}}
 \def\ba   {\begin{array}}      \def\ea   {\end{array}}
 \def\bea  {\begin{eqnarray}}   \def\eea  {\end{eqnarray}}
 \def\bean {\begin{eqnarray*}}  \def\eean {\end{eqnarray*}}

\begin{document}

\title{Cosmological Stability Bound in Massive Gravity and Bigravity}
% Cosmological Stability Bound in Massive Gravity and Bigravity
% Cosmological Solutions and Stability Bound in Massive Gravity, Bigravity
% Cosmological Solutions and Stability in Massive Gravity and Bigravity
\author{Matteo Fasiello}
\author{and Andrew J. Tolley}
\affiliation{CERCA, Department of Physics, Case Western Reserve University, 10900 Euclid Ave, Cleveland, OH 44106, USA}
%\date{\today}

%%%%%%%%%%%%%%%%%%%%%%%%%%%%%%%%%%%%%%%%%%%%%%%%%%%%%%%%%%%%%%%%%%%%%
%%%% Abstract

%Massive Gravity in four dimensions has been shown to be free of the Boulware-Deser (BD)  ghost
%both in the ADM and the \stu languages for a  specific choice of mass terms.  We show here how this is consistent with the
%helicity language.
\abstract{We give a simple derivation of a cosmological bound on the graviton mass for spatially flat FRW solutions in massive gravity with an FRW reference metric and for bigravity theories. This bound comes from the requirement that the kinetic term of the helicity zero mode of the graviton is positive definite. The bound is dependent only on the parameters in the massive gravity potential and the Hubble expansion rate for the two metrics.

We derive the decoupling limit of bigravity and FRW massive gravity, and use this to give an independent derivation of the cosmological bound. We recover our previous results that the tension between satisfying the Friedmann equation and the cosmological bound is sufficient to rule out all observationally relevant FRW solutions for massive gravity with an FRW reference metric. In contrast, in bigravity this tension is resolved due to different nature of the Vainshtein mechanism.  We find that in bigravity theories there exists an FRW solution with late-time self-acceleration for which the kinetic terms for the helicity-2, helicity-1 and helicity-0 are generically nonzero and positive making this a compelling candidate for a model of cosmic acceleration. 

We confirm that the generalized bound is saturated for the candidate partially massless (bi)gravity theories but the existence of helicity-1/helicity-0 interactions implies the absence of the conjectured partially massless symmetry for both massive gravity and bigravity. }

\maketitle

%%%%%%%%%%%%%%%%%%%%%%%%%%%%%%%%%%%%%%%%%%%%%%%%%%%%%%%%%%%%%%%%%%%%%
%%%% Introduction

\section{Introduction}

Attempts at modifying general relativity have a rich and interesting history. In recent years, this effort has been fueled  by the observation \cite{acc1,acc2} that the universe is currently accelerating: deviations from GR at large scales, a weaker gravity, or gravitationally induced self-acceleration, might account for such a dynamics.
What one hopes to gain in having an accelerated universe through modified gravity is something perhaps less mundane and certainly  more appealing than a cosmological-constant-driven acceleration \cite{prob1,prob2}. The cosmological constant problem manifests itself in all theories of modified gravity, in massive gravity (MG) it translates directly into the requirement of a small mass $m$. It has been argued that a graviton mass can be kept small in a technically natural way \cite{natural1,natural2}.

The first theory formulation of massive gravity (dRGT) in which the number of degrees of freedom nonlinearly is 5 was given in \cite{deRham:2010ik,deRham:2010kj} (see \cite{Hassan:2011hr, Hassan:2011tf, Hassan:2011ea, deRham:2011qq, deRham:2011rn,Mirbabayi:2011aa, Golovnev:2011aa, Hassan:2012qv, Deffayet:2012nr} for proofs of the absence of the Boulware-Deser ghostly 6th mode \cite{Boulware:1973my}), and this theory was subsequently generalized to arbitrary reference metric \cite{Hassan:2011tf} and to bigravity \cite{Hassan:2011ea} where both metrics are dynamical. In the following we shall be concerned with spatially flat FRW solutions in these two theories, i.e. solutions where {\it both} the dynamical and the reference metric are homogenous and isotropic.

Finding observationally relevant cosmological solutions in these theories has received considerable attention \cite{Koyama:2011yg,Chamseddine:2011bu,vonStrauss,Gumrukcuoglu:2011zh,Brihaye:2011aa,Pilo,Crisostomi,Khosravi,Vakili:2012tm,Gratia:2012wt,Kobayashi:2012fz,Volkov:2011an,Volkov:2012cf,DeFelice:2012mx,Gumrukcuoglu:2012aa,Berg,Baccetti:2012ge,Volkov:2012zb,Gumrukcuoglu:2012wt,Tasinato:2012ze,Wyman:2012iw,Sakakihara:2012iq,MartinMoruno:2013gq,Maeda:2013bha,DeFelice:2013awa,Volkov:2013roa,Tasinato:2013rza,Khosravi:2013axa,Akrami:2012vf}. In massive gravity with a Minkowski reference metric, it was shown that no homogeneous and isotropic  spatially flat solutions arise \cite{massivec}. However, it was argued in \cite{massivec} that observationally consistent inhomogenous solutions should exist (see also \cite{Comelli:2013tja,D'Amico:2012zv,Hinterbichler:2013dv,Gumrukcuoglu:2013nza,D'Amico:2013kya,DeFelice:2013tsa, Andrews:2013uca,Lin:2013aha} for different approaches). Unfortunately these inhomogenous/anisotropic solutions have been hard to find except in special cases and many of these special cases have been shown to be infinitely strongly coupled. One simple resolution is to look at massive gravity theories with an FRW reference metric \cite{Fasiello:2012rw} or for bigravity theories in which FRW solutions are allowed \cite{vonStrauss,Pilo,Volkov:2011an,Volkov:2012zb,Akrami:2012vf,Pilo,Crisostomi}. It is these classes of solution that we shall be interested in for this work. 

We will see that generically the FRW solutions in bigravity and massive gravity with an FRW reference metric (FRW massive gravity) are theoretically well defined in the sense that cosmological perturbations admit nonzero positive kinetic terms for all 5 degrees of freedom. However, for FRW massive gravity, when we go into the observationally relevant Vainshtein region \cite{Vain,Babichev:2013usa,Babichev:2013pfa}  for which the normal Friedmann equation is recovered, the helicity-0 mode becomes a ghost. This result follows because these solutions begin to violate a generalization of a cosmological bound on the mass of graviton first discovered by Higuchi \cite{Higuchi} (see also \cite{Deser,Blas,Grisa}) 

This generalization of the Higuchi bound to FRW massive gravity was derived in \cite{Fasiello:2012rw} and in this manuscript we will give two independent simpler derivations of this result. The first derivation makes use of the minisuperspace action to look for the instability, the second utilizes the massive (bi)gravity $\Lambda_3$ decoupling limit. The result obtained in \cite{Fasiello:2012rw} was that the Higuchi bound in massive gravity for an arbitrary FRW reference metric $f$ and arbitrary matter coupled to the dynamical metric $g$ is 
\be
{\tilde m}^2(H)  \ge  2 H^2
\ee
where the dressed mass $\tilde m(H)$ is given by
\be
{\tilde m}^2(H) = \frac{m^2}{2\mpl^2} \frac{H}{H_f} \left[ \beta_1 + 2 \beta_2 \frac{H}{H_f} + \beta_3 \frac{H^2}{H_f^2}\right] \, ,
\ee
$H$ is the dynamical Hubble parameter, and $H_f$ is that of the reference metric $f$. The $\beta_n$'s are the parameters in the mass potential.

The problem is that it is essentially impossible to satisfy this bound in the Vainshtein region in which the modified gravity contributions to the Friedmann equation are subdominant. This arises because both the Friedmann equation and the bound are polynomials of the same order in $H$. On the one hand, recovery of standard cosmology requires that the mass is small compared to a polynomial in $H$; on the other hand the bound requires that the mass is large compared to another a polynomial of the same order in $H$. 

In this manuscript we shall show that this tension between the Higuchi bound and the Vainshtein mechanism is generically resolved in bigravity theories, even when the Planck mass for the $f$ metric $\mpf$ is much larger than the usual Planck mass $\mpl$. This result is at first surprising, since the corrections to the action for the helicity-zero mode come suppressed by the ratio $\mpl/\mpf$. The resolution, is that even these small corrections can actually dominate the dynamics of the helicity-zero mode in the Vainshtein region because we simultaneously have $H_f/H \gg \mpf/\mpl$.

Using the two new methods of derivation, we show that in bigravity, by giving full dynamics to what is the fixed metric $f$, the bound is relaxed down to:
\bea
{\tilde m}^2(H) \left[ H^2+\frac{H_f^2 \mpl^2}{\mpf^2}\right] \geq 2H^4
\eea
where the same dressed mass enters. This is equivalent to saying that in bigravity the `dressed mass', i.e. the dynamical mass of the massive graviton, is given by $\tilde m^2(H)  (1+H_f^2 \mpl^2/(H^2 \mpf^2))$. Just as in the massive gravity case, this bound is valid for arbitrary matter content and for arbitrary FRW (i.e. it is not specific to de Sitter).
Special cases of these results have been discussed in FRW perturbation analyses of \cite{Crisostomi,Berg}), however by giving an independent derivation of them, we shall also elucidate the dynamics of the helicity-0 mode in the Vainshtein region. 

The above condition allows one to shift the burden of satisfying the bound onto the value of Hubble rate for the $f$ metric, $H_f$. Enriching massive gravity with the $f$ dynamics then proves crucial for relaxing the Higuchi bound. As expected, the massive gravity Higuchi bound is recovered in the $\mpf/\mpl \rightarrow \infty$ decoupling limit.
This bound may also be expressed in a more symmetric way which makes manifest the symmetry of the bigravity action under $H \leftrightarrow H_f$ and $\beta_n \leftrightarrow \beta_{4-n}$ and $\mpl \leftrightarrow M_g$:
\be
\frac{m^2}{2}  \left[ \beta_1 H_f^2 + 2 \beta_2 H H_f + \beta_3 {H^2}\right]  \left( \frac{H^2}{\mpl^2} +\frac{H_f^2}{\mpf^2}\right)\ \ge 2 H_f^3 H^3 \label{symm}
\ee
An intriguing, special, case of the Higuchi bound is what characterizes the so-called \textit{partially massless} (PM) theory of massive gravity \cite{Deser:1983mm,Deser:2001pe,Deser:2001us,Deser:2001xr,Deser:2004ji,Zinoviev:2001dt,Zinoviev:2006im,Deser:2006zx,Joung:2012hz,deRham:2012kf,Deser:2013xb,Deser:2013uy,Hassan:2012gz,deRham:2013wv}. There the Higuchi inequality is actually saturated for all values of $H$ regardless of the matter coupling to the $g$ metric and the Higuchi-Vainshtein tension is resolved in that massive gravity parameters $\beta_n$ act in such a way that one of the two scale factors becomes pure gauge. 

The special parameter values for the potentially partially massless theory were first identified in massive gravity \cite{deRham:2012kf}  and they are the same parameters that have also been used in bigravity \cite{Hassan:2012gz}. They correspond to $\beta_1=\beta_3 =0$, $\beta_0 = \frac{3}{2} \mpl^4/\mpf^2$, $\beta_2 = \mpl^2$, $\beta_4 = \frac{3}{2} \mpf^2$. 
For this special case, the bound becomes:
\be
{m^2}  \mpl^2  \left( \frac{H^2}{\mpl^2} +\frac{H_f^2}{\mpf^2}\right)\ \ge 2 H_f^2 H^2,
\ee
but, using the Friedmann equation for the $f$ metric, we have in this specific case that
\be
H_f^2 \mpf^2 =\frac{1}{2} m^2 \mpl^2  \(\frac{H_f^2}{H^2} + \frac{\mpf^2}{\mpl^2} \). 
\ee
Substituting in we find that the bound becomes an identity regardless of the value of $H$ and $H_f$. This represents a nontrivial check on the consistency of the derived bound.

We shall give two independent derivations of the new bounds. The first is the simplest and utilizes the minisuperspace action alone. We present the reasoning here since it may be useful in similar analyses for more general theories. The idea is that since the bound comes from requiring the absence of ghosts for the helicity-0 mode, and the helicity-0 mode being a scalar already enters into the minisuperspace Lagrangian\footnote{Minisuperspace is the truncation of a gravity theory to the case where the metric and all fields are functions of time alone. In other words it the compactification down to $0+1$ dimensions.}, it should be possible to identify the sign of the kinetic term for the helicity-zero mode by analyzing the minisuperspace action alone. Performing this analysis for massive gravity we recover our result of  \cite{Fasiello:2012rw} which was obtained by a much more laborious Hamiltonian analysis. This method easily generalizes to the bigravity case and for generic matter couplings.   

The second derivation utilizes the $\Lambda_3$ decoupling limit of massive gravity and the $\Lambda_3$ decoupling limit of bigravity which we derive here in full, including the vector degrees of freedom following the approach of \cite{Ondo:2013wka} (see also \cite{Gabadadze:2013ria}). It was not guaranteed that this approach would agree with the exact answer since in principle it is possible that the exact bound contains terms which vanish in the decoupling limit. However, this is not the case. Taking the bigravity bound, and scaling $m \rightarrow 0$, $\mpl \rightarrow \infty$ keeping the ratios $H/m$, $\mpl/\mpf$, $H/H_f$ and the parameters $\hat \beta_n = \beta_n/\mpl^2$ fixed, we find that the bound remains the same since it can be expressed entirely in terms of the fixed ratios
\be
\frac{1}{2}  \left[ \hat \beta_1 \frac{H_f^2}{m^2} + 2 \hat \beta_2 \frac{H}{m} \frac{H_f}{m} + \hat \beta_3 {\frac{H^2}{m^2}}\right]  \left( \frac{H^2}{m^2} +\frac{\mpl^2}{\mpf^2} \frac{H_f^2}{m^2}\right)\ \ge 2 \frac{H_f^3}{m^3} \frac{H^3}{m^3}  \, .
\ee
Since this is the limit used in deriving the massive gravity and bigravity $\Lambda_3$ decoupling limit, it follows that the cosmological bound can be entirely determined by the decoupling limit. The decoupling limit also allows us to go beyond the linearized analysis of \cite{Crisostomi} since it captures the leading nonlinear interactions that will be relevant in cosmological perturbation theory. 

Finally, our derivation of the $\Lambda_3$ decoupling limit for bigravity leads to a new surprise, a dual formulation of Galileons. In massive gravity, the `Galileon' \cite{galileon1} arises as the helicity-0 mode in the map between the coordinate systems of the $g$ metric ($x^a$) and the $f$ metric ($\Phi^a$) \cite{deRham:2010ik}, 
namely
\be
\Phi^a(x) = x^a + \frac{1}{\La^3_3} \eta^{ab} \frac{\partial \pi(x)}{\partial x^b}  \, 
\ee
we will prove that this transformation admits an inverse of the form
\be
x^a(\Phi) = \Phi^a +\frac{1}{\La^3_3} \eta^{ab}  \frac{\partial \rho(\Phi)}{\partial \Phi^b}  \, .
\ee
$\rho(x)$ is the dual Galileon field which is related to $\pi(x)$ by a nonlocal field redefinition. It is the Galileon field viewed from the perspective of the $f$ metric. The decoupling limit of bigravity must be symmetric with respect to interchange of the two metrics, and hence it is symmetric with respect to the interchange of $\rho$ and $\pi$. We shall discuss this duality in more detail elsewhere \cite{duality}.

The paper is organized as follows. In {{\it Section}} \ref{MGderiavtion} we introduce the dRGT theory of massive gravity and determine the bound using the minisuperspace approach. In {\it Section} \ref{Gen} we generalize the stability bound to bigravity using the minisuperspace method and find that it is much easier to satisfy than in massive gravity. In {\it Section}  \ref{Dec} we derive the $\Lambda_3$ decoupling limit of bigravity and by extension that of massive gravity on a generic reference metric.  In {\it Section} \ref{BiDec} we use the bigravity decoupling limit to give an independent derivation of the cosmological bounds. Finally, we discuss the conjectured partially massless gravity and bigravity in {\it Section} \ref{PM}, establish that they saturate the bound, and discuss their validity as candidate partially massless theories.  In the {\it Appendix} \ref{Appendix} we give the details of the bound for fluctuations around de Sitter. 

\section{Deriving the bound in Massive Gravity on FRW}

\label{MGderiavtion}

We now give a new derivation of the cosmological bound in massive gravity with a spatially flat FRW reference metric. This confirms the result obtained in \cite{Fasiello:2012rw} and we present it here as a warm up to the analogous bigravity calculation in the next section.

The theory of massive gravity defined on an arbitrary reference metric $f_{\mu\nu}$ \cite{Hassan:2011tf} is just a straightforward generalization of the theory proposed in \cite{deRham:2010kj}. The Lagrangian takes the form of Einstein gravity with matter plus a potential that is a scalar function of the two metrics
\be
\mathcal{L}=\frac{\mpl^2}{2}\sqrt{-g}\left(R+{2 m^2}\mathcal{U}(\cal K)\right)+ \mathcal{L}_M\,. \label{nomatter}
\end{equation}
The most general potential $\mathcal{U}$  that has no ghosts \cite{deRham:2010kj} is built out of characteristic (symmetric) polynomials of the eigenvalues of the tensor
\be
\mathcal{K}^{\mu}_{\nu}(g,f)=\delta^{\mu}_{\nu}-\sqrt{g^{\mu \alpha}f_{\alpha \nu}} ,
\ee
so that
\begin{equation}
\label{eq:fullU}
\mathcal{U}({\cal K})=\mathcal{U}_2+\alpha_3 \, \mathcal{U}_3+\alpha_4 \, \mathcal{U}_4,
\end{equation}
where the $\alpha_n$ are free parameters, and
\begin{eqnarray}
\mathcal{U}_2&=& \frac{1}{2!}\left( [\mathcal{K}]^2-[\mathcal{K}^2]\right),\\
\mathcal{U}_3&=&\frac{1}{3!}\left( [\mathcal{K}]^3-3[\mathcal{K}][\mathcal{K}^2]+2[\mathcal{K}^3]\right),\\
\mathcal{U}_4&=&\frac{1}{4!}\left(  [\mathcal{K}]^4-6[\mathcal{K}^2][\mathcal{K}]^2+8[\mathcal{K}^3][\mathcal{K}]+3[\mathcal{K}^2]^2-6[\mathcal{K}^4]\right) \,,
\end{eqnarray}
where $[\ldots]$ represents the trace of a tensor with respect to the metric $g\mn$. The $\mathcal{U}_n$ are the symmetric polynomials which are generally defined in $D$ dimensions by the determinant relation
\be
{\rm Det}[1 +\lambda {\cal K}] = \sum_{n=0}^D \lambda^n \, {\cal U}_n(\cal K) \, .
\ee
so that $\mathcal{U}_0=1$ and $\mathcal{U}_1= [\mathcal{K}]$.
The Lagrangian may also be written in the form
\be
\mathcal{L}=\frac{1}{2}\sqrt{-\, {}^{}\!g}\left[\, \mpl^2 \, R-{ m^2}\sum_{n=0}^4 \beta_n \, {\mathcal{U}}_n(X) \right]+ \mathcal{L}_M\,, \label{nomatter}
\end{equation}
where $X^{\mu}_{\nu} =\sqrt{g^{\mu \alpha}f_{\alpha \nu}}$. The relationship between the $\beta_n$ coefficients and the $\alpha_n$ is given in Eq.(\ref{conversion}). The mass term is invariant under the simultaneous interchange $g \leftrightarrow f$ and $\beta_n \leftrightarrow \beta_{4-n}$. 

As we shall see in detail, the generalization of the Higuchi bound can already be seen at the level of the minisuperspace action, in particular in the representation of massive gravity which includes the \stu fields. The \stu fields for diffeomorphisms are introduced by replacing the reference metric $f_{\mu\nu}(x)$ by its representation in an arbitrary coordinate system $\Phi^A(x)$ 
\be
ds_f^2 = f_{AB}(\Phi^C) \partial_{\mu} \Phi^A\partial_{\nu} \Phi^B dx^{\mu} dx^{\nu} \, .
\ee
The $\Phi^A(x)$ are the \stu fields (or Goldstone modes for the broken diffemorphisms) and essentially encode the additional degrees of freedom that a massive graviton has over a massless one. These additional degrees of freedom are,  in the high energy limit, decomposable into 2 helicity-1 modes and 1 helicity-0 mode. Explicitly, writing $\Phi^A(x) =B^A(x)/(m \mpl) + \partial^A \pi(x)/\La_3^3$, then $B^a$ has the interpretation of the helicity-1 mode in the high energy limit, and $\pi$ is the additional helicity-0 mode. Since the bound comes from identifying the sign of the kinetic term of the helicity-0 mode, it is essentially enough to keep track of the $\Phi^0 \sim \partial_t \pi$ term (we may for instance choose a gauge for which $B^0=0$ to aid in this identification).

However as is well known, in massive gravity, either part or all of the kinetic term for the helicity-0 mode actually comes from a mixing of $\Phi^0$ and the metric $g_{\mu \nu}$. In perturbations, this shows up in the fact that a scalar part of $g_{\mu\nu}$ couples to $\Phi^0$ and hence $\dot \pi$. In the minisuperspace limit, the only scalar part of the metric is the scale factor itself $a(t)$\footnote{Although the lapse $N(t)$ is also a scalar, it is a Lagrange multiplier for a constraint and drops out of the action if the constraint is solved.}. Thus we can identify the sign of the kinetic term for perturbations around a background that identifies kinetic term for $\delta a$, the mixing between $\delta a$ and $\delta \Phi^0$, and any independent kinetic term for the helicity-0 which enters as ${\delta \Phi^0}^2$. This method is significantly simpler than our previous complete calculation \cite{Fasiello:2012rw} and leads quite quickly to the same result.

\subsection{Minisuperspace Derivation}

Thus, we begin with the action for minisuperspace for the dRGT model on a given FRW background written in an arbitrary gauge by means of the introduction of a Stueckelberg field for the broken time diffeomorphisms. The reference metric is given by
\be
ds_f^2 = - \dot{\phi}^2 dt^2 + b(\phi)^2 d \vec{x}^2 \, ,
\ee
where $\phi= \Phi^0$ is the \stu field which will keep track of the helicity-0 mode kinetic term $\delta \phi \sim \partial_t \delta \pi$. In this section, for the sake of simplicity, we will limit ourselves to the case of a de Sitter reference metric.  The more general case is dealt with in the bigravity setup of \textit{Section 3}.
The dynamical metric $g$ is similarly expressed as
\be
ds_g^2 = - N(t)^2 dt^2 + a(t)^2 d \vec{x}^2 \, .
\ee
The dRGT square root combination takes now the form 
\be
\sqrt{g^{-1}f} = 
\left(\begin{array}{cc}
\frac{\dot{\phi}}{N} & 0_j  \\
0_i & \frac{b(\phi)}{a} \delta_{ij}  \\
\end{array} \right) \, ,
\ee
and we have chosen the square root with the positive sign (see \cite{Deffayet:2012zc,Gratia:2013gka} for a discussion on this). It follows that the minisuperspace action for spatially flat cosmological solutions takes the form
\be
S = V_3 \int \d t N a^3 \left[ -3 \mpl^2 \left( \frac{\dot{a}^2}{N^2 a^2}\right)- \frac{\dot{\phi}}{N} \sum_{n=0}^3 A_n \left(\frac{b(\phi)}{a} \right)^n- \sum_{n=0}^3 B_n \left(\frac{b(\phi)}{a} \right)^n - \rho(a) \right],
\ee
where $V_3 = \int \d^3 x$ is the volume factor and $\rho(a)$ is the matter density. 

The coefficients $A_n$ and $B_n$ are subject to the relation $A_n(3-n)=B_{n+1}(n+1)$ and in terms of the $\alpha_n$ and $\beta_n$, these coefficients are given by
\bea
\label{conversion}
&&B_0 = m^2 \mpl^2 (-6-4 \alpha_3 - \alpha_4) = \frac{1}{2} m^2 \beta_0 \nonumber  \\
&&B_1 = 3m^2 \mpl^2(3+3 \alpha_3+\alpha_4 ) = \frac{3}{2}m^2 \beta_1 \nonumber  \\
&&B_2 = 3  m^2 \mpl^2 (-1-2 \alpha_3-\alpha_4)=\frac{3}{2}m^2  \beta_2 \nonumber  \\
&&B_3 = m^2 \mpl^2(\alpha_4+\alpha_3)=\frac{1}{2}m^2  \beta_3 \label{conv} \, ,
\eea
\bea
B_n=\frac{3m^2 \beta_n}{(3-n)! n!}\,; \qquad A_n=\frac{3 m^2 \beta_{n+1}}{(3-n)! n!} \, .
\eea
In massive gravity, we can set $A_3=\beta_4=0$ since this term is a total derivative. In bigravity this term enters as the cosmological constant sourcing the $f$ metric and thus should in general be maintained.

The Friedmann equation, which can be obtained by varying the action with respect to $N$, is:
\be
 H^2 = \frac{1}{3M_{P}^2} \left( \rho + \rho_{m.g.} \right),
 \label{fried1}
\ee
where the extra `dark energy' contribution from the massive gravity action is
\be
 \rho_{m.g.} = \sum_{n=0}^3 B_n \left(\frac{b(\phi)}{a} \right)^n .
\ee
Varying the action with respect to $\phi$ imposes, as a consequence of the special relation between the coefficients $A_n$ and $B_n$, the constraint
\be
\left( \sum_{n=0}^2  \frac{\hat \beta_{n+1}}{(2-n)! n!} \left( \frac{b}{a} \right)^{n+1}   \right) \left( \frac{H}{b} - \frac{H_f}{a} \right) =0 \, ,
\ee
where 
\be
H_f = \frac{\dot{b}}{\dot{\phi} b}=\frac{b_{,\phi}}{b} .
\label{fried2}
\ee
At first sight there appear to be two branches of solutions to this equation. However, if we choose the term in the first parenthesis to vanish, the kinetic term for the helicity-1 mode vanishes and all such solutions are infinitely strongly coupled. Thus the only acceptable solution is 
\be
\frac{b}{a}=\frac{H}{H_f} \, .
\ee 
From now on, we will work in the normal branch for which this relation is true (branch 2 in the language of \cite{Crisostomi}). This allows us to write the contribution to the Friedmann equation in the form 
\be
 \rho_{m.g.} = \sum_{n=0}^3 B_n \left(\frac{H}{H_f} \right)^n = \sum_{n=0}^3  \frac{3m^2 \beta_n}{(3-n)! n!} \left(\frac{H}{H_f} \right)^n \, .
\ee \\
To determine the kinetic term for the helicity zero mode we will utilize some of the properties of the minisuperspace action. 
Starting from the action 
\be
S = V_3 \int \d t N a^3 \left[ -3 \mpl^2 \left( \frac{\dot{a}^2}{N^2 a^2}  \right)- \frac{\dot{\phi}}{N} \sum_{n=0}^3 A_n \left(\frac{b(\phi)}{a} \right)^n- \sum_{n=0}^3 B_n \left(\frac{b(\phi)}{a} \right)^n - \rho(a) \right],
\ee
we now perturb it to second order in  $\delta a, \delta N, \delta \phi$. As $\delta N$ appears only algebraically in the action, it can be integrated out. After these steps the action takes the form: 
\bea
S^{(2)} &=&  V_3 \int \d t \Bigg[ \delta \phi^2 \left(   -  \frac{b_{\phi}^2}{2} \sum_{n=0}^{3}B_n n (n-1) \frac{b^{n-2}}{a^{n-3}}       +\frac{\dot a \,b_{\phi}}{2} \sum_{n=0}^{3}A_n (3-n) n\frac{b^{n-1}}{a^{n-2}}    \right.\\   &&
\left.    +\frac{b_{\phi}}{12 \mpl^2 a \dot{a}^2} \left(\sum_{n=0}^{3} n B_n \frac{b^{n-1}}{a^{n-3}} \right)^2  -\frac{H_f b_{\phi}}{2}\sum_{n=0}^{3}  B_n n \frac{b^{n-1}}{a^{n-3}}        \right)                  +\delta\phi \, \delta{\dot a}^2 \left(.. \right) +\delta a^2 \left(.. \right)    \Bigg],  \nn  \label{fluc}
\eea
where $b_{\phi}=db(\phi)/d\phi$. We see that, with the sign of the kinetic term of the helicity zero mode in mind, we need only worry about the $\delta \phi^2$ coefficient (it is $\delta \phi$ that is carrying the helicity zero mode, $\delta \phi\sim \delta \dot \pi$ ) in the  above $S^{(2)}$: integrating out $\delta N$ removes any ${\delta \dot a}^2$-proportional term in the action for fluctuations so that the last two terms in (\ref{fluc}) need not be specified, i.e. we do not need to diagonalize.

One can now make use of the following set of relations:
the background equations of motion imply  $ b_{\phi}-\dot a=0$ and we have $A_{n}(3-n)=B_{n+1}(n+1)$ , so that
\be
\sum_{n} n B_n \left(\frac{b}{a}\right)^n=\frac{b}{a} \sum_{n} A_{n} (3-n)  \left(\frac{b}{a}\right)^n.
\ee
Using these in Eq.(\ref{fluc}) leads to a further simplified action, whose helicity-0 kinetic term reads
\be
S_{(2)}\big|_{{\rm kinetic}} = V_3 \int dt\,\, \delta\phi^2\,\, \frac{3}{4} \frac{\mpl^2 a^3  }{H^2} \frac{b_{\phi}^2}{b^2}\  \Big[\tilde{m}^2 \left( \tilde{m}^2 -2H^2  \right ) \Big] ,
\ee
and we have defined the dressed mass parameter as
\bea
\tilde{m}^2(H) &=& \frac{1}{3{\mpl^2}}  \frac{b}{a} \sum_{n} A_{n} (3-n) \left( \frac{b}{a} \right)^n \nn
\\&=&  \frac{m^2}{2\mpl^2} \frac{H}{H_f} \left[ \beta_1 + 2 \beta_2 \frac{H}{H_f} + \beta_3 \frac{H^2}{H_f^2}\right] \, .  \label{dress}
\eea 
Thus, the Higuchi bound, which amounts to the statement that the kinetic term for the helicity zero mode $\pi$ is positive, reads 
\be
\tilde{m}^2(H) (\tilde{m}^2(H)-2H^2) \ge  0\, .
\ee
Upon neglecting the branch of solutions $\tilde{m}^2(H)<0$, which inevitably gives a ghost in the vector sector, the bound translates into the more familiar statement:
\be
\tilde{m}^2(H) \ge 2 H^2, \label{bmg}
\ee
where $\tilde{m}^2(H)$ is the dressed mass, a generalization of the bare $m$. This is the generalization of the Higuchi bound first derived in \cite{Higuchi}.
Once more, let us stress that this result follows regardless of the matter content and is independent of $\dot{H}$. This result is consistent with that derived in \cite{Fasiello:2012rw} by an entirely different analysis. We now have a more clear explanation of how it arises and, as we shall see, a quicker route to its extensions and applications.

\subsection{Higuchi versus Vainshtein tension}

We take now a moment to briefly reproduce the reasoning that lead in \cite{Fasiello:2012rw} to the conclusion that FRW on FRW solutions in massive gravity are ruled out. The crucial point is that the bound in Eq.~(\ref{bmg}) must be complemented with a condition on $m$ derived from consistency of the expansion history with observations. Departures from the General Relativity (GR) expansion history are negligible at large redshifts and hence large $H$; thus the $H$ dependent modifications to the normal Friedmann equation must be small, at least for most of the history of the universe. In massive gravity, continuity with GR is achieved through the Vainshtein mechanism and it is for this reason that we refer to these two opposing requirements as the Higuchi-Vainshtein tension.

In order for the $H$ dependent $m^2$ contribution in the Friedmann equation (\ref{fried1}) to be subleading (Vainshtein regime) we require:
\bea
\frac{m^2}{2\mpl^2} \left[ 3\beta_1\frac{H}{H_f} +3\beta_2\frac{H^2}{H_f^2} +\beta_3\frac{H^3}{H_f^3} \right]  \ll 3H^2 \, .
\eea
This is immediately at odds with the stability (Higuchi) condition. We can see why by combining the two inequalities into the single statement
\bea
\left[\frac{3}{2} \beta_1\frac{H}{H_f}+3 \beta_2\frac{H^2}{H_f^2}+\frac{3}{2}\beta_3 \frac{H^3}{H_f^3}   \right]       \gg \left[ 3\beta_1\frac{H}{H_f} +3\beta_2\frac{H^2}{H_f^2} +\beta_3\frac{H^3}{H_f^3} \right]  \,  \label{together}.
\eea
 Eq.~(\ref{together}) is essentially impossible to satisfy. If the $\beta_1$ term dominates the inequality is violated, if the $\beta_2$ dominates it is saturated and if $\beta_3$ dominates it is only just satisfied since $3/2>1$.  However, this is not enough, there needs to be a large hierarchy between the two sides otherwise there will be ${\cal O}(1) $ modifications to the Friedmann equation \cite{Fasiello:2012rw}. Considering that the combined inequality must hold over different cosmological epochs, one comes to the realization that  there is no room in the parameter space of FRW massive gravity for it to simultaneously satisfy the requirements of stability and consistency with observations. 

We stress that this does not rule out massive gravity as a theory of current cosmological expansion. There are many paths one can follow in the search for solutions to massive gravity that, if not exactly FRW, at least  resemble FRW in appropriate regions, see e.g. \cite{massivec}. We choose here to instead demand exactly FRW solutions and move on to give full dynamics to the reference metric $f$, thus entering into the realm of bigravity theories.

\section{Generalizing the Bound to Bigravity}

\label{Gen}

The action for bigravity models which are free from the Boulware-Deser ghost is a simple extension of that for massive gravity  \cite{Hassan:2011ea}
\be
S=\int \d^4 x \, \frac{1}{2}\left[\, \mpl^2 \sqrt{-g} \, R[g]+\mpf^2 \sqrt{-f} \, R[f]-{ m^2}\sum_{n=0}^4 \beta_n \, {\mathcal{U}}_n(X) \right]+ \mathcal{L}_M\,. \label{nomatter} 
\end{equation}
It is straightforward to generalize the previous argument to the case of bigravity. We denote the now dynamical second metric $f$ as 
\be
ds_f^2= - \tilde{N}^2 dt^2 + b^2 d \vec{x}^2.
\ee
The minisuperspace action is now 
\bea
 S &=& V_3 \int \d t a^3 \left[ -3 \mpl^2 \left( \frac{\dot{a}^2}{N a^2} \right)- \tilde{N} \sum_{n=0}^3 A_n \left(\frac{b}{a} \right)^n- N \sum_{n=0}^3 B_n \left(\frac{b}{a} \right)^n - N \rho(a)  \right.   \nonumber  \\
 &+&  V_3 \int \d t b^3  \left. - \tilde{N} \tilde{\rho}(b)  -3 \mpf^2 \left( \frac{\dot{b}^2}{\tilde{N} b^2} \right) \right]
\eea
The Friedmann equation for the scale factor $a$, which we assume to correspond to the metric with which our matter is coupled, is given by
\be
H^2 = \frac{1}{3 \mpl^2} \left( \rho(a) + \rho_{\rm bigravity} \right) \, ,
\ee
where 
\be
\rho_{\rm bigravity} = \sum_{n=0}^3 B_n \left(\frac{b}{a} \right)^n  .
\ee
In the following we will assume that the only matter sourcing the second metric is a cosmological constant and, as such, $\tilde{\rho}(b)$ can be absorbed into a definition of $A_3$. In this case the Friedmann equation for the second metric is simply 
\be
H_f^2 = \frac{1}{3 \mpf^2} \Bigg[ \sum_{n=0}^3 A_n \left(\frac{b}{a} \right)^{(n-3)} \Bigg] 
\ee
For convenience let us denote 
\be
{\cal A}(\chi) = \sum_{n=0}^3 A_n \left(\frac{b}{a} \right)^n =\sum_{n=0}^3 A_n e^{n \chi} \, ,\quad \quad
{\cal B}(\chi)=\sum_{n=0}^3 B_n \left(\frac{b}{a} \right)^n=\sum_{n=0}^3 B_n e^{n \chi} \, ,
\ee
where we have utilized the following change of variables $b/a=e^{\chi}$. Let us also employ a simple yet generic matter Lagrangian coupled only with the $g$ metric:
\bea
\mathcal{L}_M = -\frac{1}{2}\,g^{\mu\nu}\partial_{\mu}\phi\partial_{\nu}\phi-V(\phi)\,,
\eea
so that we add to the minisuperspace action the term
\be
S_M = V_3 \int \d t N a^3 \left( \frac{1}{2} {\dot \phi}^2 - V(\phi) \right)  \, .
\ee
We shall see that all details of the precise nature of the matter Lagrangian will drop out of the final answer.  We now pass to the canonical phase space formulation by defining the canonically conjugate momenta
\be
p_a = -6 \mpl^2 \frac{\dot{a}}{N}a \, , \quad p_{\phi}= a^3 \frac{\dot{\phi}}{N} \, , \quad p_b=-6 \mpf^2 \frac{\dot{b}}{{\tilde N}} b \, .
\ee
As a consequence, the canonical action can be written as
\be
S =  V_3\int \d t \Bigg[ \, p_a \dot{a} + p_b \dot{b}+ p_{\phi} \dot{\phi}  - N \left(-\frac{1}{2} \frac{p_a^2}{6 \mpl^2 a}+a^3 {\cal B}+ \frac{p_{\phi}^2}{2a^3}+a^3 V \right) -\tilde{N} \left(-\frac{1}{2} \frac{p_b^2}{6 \mpf^2 b}+a^3 {\cal A} \right)  \Bigg] .
\ee
Varying with respect to $\tilde{N}$ and $N$ imposes the analogues of the Hamiltonian constraints which can be solved to remove $p_b$ (we will choose the negative solution corresponding to an expanding geometry) 
\be
p_a = -\sqrt{12} a^2 \mpl \sqrt{{\cal B}+ \frac{p_{\phi}^2}{2 a^6}+V} \, , \quad \,  p_b = -\sqrt{12} \mpf \sqrt{b {\cal A}a ^3},
\ee
and, using $b=a\,e^{\chi}$, we obtain 
\be
S = V_3 \int \d t \, \Bigg[ p_{\phi} \dot{\phi}-\sqrt{12} a^2 \mpl \dot{a} \sqrt{{\cal B}+ \frac{p_{\phi}^2}{2 a^6}+V}      - \sqrt{12} a^2 (\dot{a}+\dot{\chi} a ) \mpf \sqrt{e^{3 \chi} {\cal A}}\,  \Bigg]  , 
\ee
or, after solving for $p_{\phi}$ and integrating out,
\be
S =  V_3\int \d t \Bigg[ \, -\sqrt{12} a^2 \mpl \dot{a} \sqrt{{\cal B}+ V }\sqrt{1-\frac{a^2 \dot{\phi}^2}{6 \mpl^2}}  - \sqrt{12} a^2 \mpf \sqrt{e^{3 \chi} {\cal A} }(\dot{a}+\dot{\chi} a )  \Bigg]  \, .
\ee
As it stands, this action is time reparametrization invariant, and we may utilize this fact to choose the gauge $a=t$, so that
\be
S =  V_3\int \d t \, \Bigg[ -\sqrt{12} t^2 \mpl \sqrt{{\cal B}+ V }\sqrt{1-\frac{t^2 \dot{\phi}^2}{6 \mpl^2}}  - \sqrt{12} t^2 (1+\dot{\chi} t )  \mpf \sqrt{e^{3 \chi} {\cal A} }    \Bigg] \, . 
\ee
We may now perturb this action, $\phi \rightarrow \phi + \delta \phi$ and $\chi \rightarrow \chi + \delta \chi$,  to second order so as to obtain the quadratic action for perturbations around a given background. 
We also note that the contribution from fluctuations of the potential $V(\phi)$ will at most contribute a mixing term $\sim \phi\, \chi =  \phi \, \dot \pi$;  it does not therefore add up to the kinetic term whose sign we are after and can thus be ignored in what follows\footnote{Note that the variable $\phi$ here is the scalar in the matter Lagrangian, not to be confused with the Stueckelberg field $\phi$ in the massive gravity analysis of the previous section}.

Up to an unimportant positive overall factor, the quadratic  fluctuation action reads:
\bea
S_{(2)} \propto  V_3 \int \d t \, \tilde{m}^2 \left( \frac{H_0^2}{4} \frac{\mpl^2 \tilde{m}^2}{H^5 \mpf^2} - \frac{1}{2 H} + \frac{\tilde{m}^2}{4H^3(1-g^2)}    \right) \, \delta \chi^2 + g \frac{ \tilde{m}^2}{H(1-g^2)}\, \delta \chi \delta g +\frac{ H }{1-g^2} \, \delta g^2 \nonumber \\
\eea
where we have defined the dimensionless variable $g$ out of $\dot \phi$ as $g^2\equiv t^2 \dot \phi^2/6 \mpl^2 $. We have also employed Eq.~(\ref{fried1}) and (\ref{fried2}) to simplify the coefficients: this provides a more compact expression and is of course allowed as coefficients are purely background quantities. 

Since $\delta \chi$ carries the information on $\pi$, the helicity zero mode, $\delta \chi \propto \delta \dot{\pi}$,  we proceed to diagonalize the above expression so as to read off the $\dot\pi^2$ coefficient. If we posit 
\be
\delta g= \delta\tilde{g}- \frac{g\, \tilde{m}^2}{2H^2}\,\, \delta\chi ,
\ee
in the new variables the mixing term vanishes and, requiring the $\delta \chi^2$ coefficient be positive, amounts to: 
\be
\tilde{m}^2 \left(H^2+ \frac{\mpl^2 H_f^2}{\mpf^2} \right) -2H^4 \geq 0  \,\,\,  , \label{main}
\ee
%54
that is, the generalized Higuchi bound.  \\

We stress again that simple algebraic manipulations coupled with the properties of the $A_n,B_n$ (or, equivalently, the $\beta_n$) coefficients allow Eq.~(\ref{main}) to be expressed in its most symmetric and pleasing form, Eq.(\ref{symm}), which we reproduce here:
\bea
\frac{m^2}{2}  \Big[ \beta_1 H_f^2 + 2 \beta_2 H H_f + \beta_3 {H^2}\Big]  \left( \frac{H^2}{\mpl^2} +\frac{H_f^2}{\mpf^2}\right)\ \ge 2 H_f^3 H^3 \label{symm2} \,.
\eea
Let us reiterate that this bound was derived for fairly generic matter content, but matter itself enters into the final expression only indirectly through $H$. Thus the bound is entirely one on the Hubble rate, regardless of the matter content of the universe, and may thus be universally applied at all cosmological epochs.

\subsection{Higuchi versus Vainshtein Resolution}

As stressed in \cite{Fasiello:2012rw}, one should not simply declare cosmological solutions viable by merely looking for regions of the parameter space where Eq.~(\ref{symm2}) is satisfied. Rather, we ought to perform a joint analysis of the stability condition and the requirement stemming from observations that, for most of the history of the universe, the Friedmann equation is well approximated by its GR prediction. In essence this requires us to be in the Vainshtein regime for which the cosmological effects of the helicity-0 mode are negligible and are suppressed by powers of $m$. Unfortunately these two requirements oppose each other, thereby creating the Higuchi-\textit{vs}-Vainshtein tension that effectively rules out the viability of spatially flat FRW solutions in FRW massive gravity. \\

Let us see how bigravity resolves the Higuchi-Vainshtein tension; the ingredients are the generalized bound and the Friedmann equations, which we report below: 
\bea
H^2 = \frac{1}{3 \mpl^2} \Bigg[ \rho(a) +  \sum_{n=0}^3 \frac{3m^2 \beta_n}{(3-n)! n!}\left(\frac{H}{H_f} \right)^n \Bigg] \, ;  \,\,\, H_f^2 = \frac{1}{3 \mpf^2} \Bigg[ \sum_{n=0}^3      
\frac{3\beta_{n+1}}{(3-n)! n!}  \left(\frac{H}{H_f} \right)^{(n-3)} \Bigg].  \label{frieds} \nonumber \\
\eea
Now, the inequality in Eq.~(\ref{main}) is certainly satisfied, without requiring much on the part of $\tilde{m}$, if we posit $H_f\gg H$. In particular if we assume that $\beta_1$ is nonzero, as is generically the case, then the second relation in (\ref{frieds}) and the $\tilde{m}$ definition (see \ref{dress}) are, at leading order in $H/H_f$, respectively:
\bea
 3 \mpf^2 H_f^2\sim m^2 \frac{\beta_1}{2} \frac{H_f^3}{H^3}\,;\quad  \tilde{m}^2 \sim m^2  \frac{\beta_1}{2\mpl^2} \frac{H}{ H_f} .
\eea
From here, we take the ratio of the two expressions, solve for $H_f$ and plug it back into Eq.~(\ref{main}), to obtain:
\bea
H_f \sim \sqrt{3}\frac{\mpf H^2}{\mpl \,\tilde{m}};\qquad   \qquad {\rm Stability \,bound}\Big|_{H_f\gg H}: \,\,3H^4 \gtrsim 2H^4  \,\,\,, 
\eea
which is, evidently, consistent with the initial assumption.  

As for the first Friedmann equation in (\ref{frieds}), we see that, by itself,  it does not put too much strain on $\tilde{m}$. The fact is that the dressed mass is actually allowed to be small because precisely the lower bound on $\tilde m$ that in the massive gravity case comes from the stability condition, is now taken care of by our being in the $H_f\gg H$ region. \\

We have thus shown  that there exists  a large region in the parameter space of the bigravity theory, $H \ll H_f$, which does relax the stability bound in a way which is consistent with observations. The stability-\textit{vs}-observation conflict is therefore resolved. This resolution requires that in the cosmological bound, the term $H_f^2/\mpf^2$ dominates over $H^2/\mpl^2$. It is clear that in the FRW massive gravity limit, this could never be satisfied since that limit requires us to take $\mpf \rightarrow \infty$ for finite $\mpl$ and finite $H_f$ and so inevitably the former is subdominant to the latter. Thus it is precisely the naively $\mpf$ suppressed bigravity interactions, which are crucial in resolving the conflict. 

\subsection{Ghost-free Self-Accelerating Cosmology}
\label{cosmology}

Having established the regime we should be in to satisfy the bound, we can infer the effective Friedmann equation by solving 
\be
 3 \mpf^2 H_f^2\sim m^2 \frac{\beta_1}{2} \frac{H_f^3}{H^3}
\ee
to give
\be
H_f \sim \frac{6 \mpf^2 H^3}{m^2 \beta_1}
\ee
and then substituting back into the $H$ Friedmann equation to give
\be
H^2 \sim \frac{1}{3 \mpl^2} \Bigg[ \rho(a) + \frac{3 m^2 \beta_1}{2} \left( \frac{H}{H_f} \right) \Bigg] =  \frac{1}{3 \mpl^2}  \Bigg[ \rho(a) + \frac{m^4 \beta_1^2}{4 \mpf^2} \left( \frac{1}{H^2} \right) \Bigg]  \, .
\ee
This gives a self-accelerating cosmology in which asymptotically in the future the Hubble parameter tends to a constant de Sitter value of magnitude
\be
H_{\infty} = \frac{1}{12^{1/4}} m  \frac{\beta_1^{1/2}}{\sqrt{\mpl \mpf}}.
\ee
Thus we see that quite in general the condition $H_f/H \gtrsim \mpf /\mpl$, which is implicitly assumed in having $\beta_1$ dominate, is essentially satisfied even in the asymptotics \footnote{Note that in the asymptotics it is acceptable to have $H_f/H \sim \mpf /\mpl$, this is because, even if it implies via (\ref{main}) that $\tilde{m}^2\sim H^2$, the latter is not an issue since at late times we desire order unity departures from GR to account for cosmic acceleration.}:  
\be
\frac{H_f}{H} \sim \frac{6 \mpf^2 H^2 }{m^2 \beta_1} \sim \frac{6 \mpf}{\sqrt{12} \mpl}
\ee
Also note that, since $H/H_f$ scales as $1/H^2$,  it becomes easier and easier to satisfy the condition as we go back in time. 

To make this discussion more precise, we can specify to the model with $\beta_3=0$, $\beta_2=0$ and $\beta_1 = 2 \mpl^2$ ($\alpha_3=-1$, $\alpha_4=1$) . In this case the $H$ Friedmann equation is exactly
\be
H^2 = \frac{1}{3 \mpl^2}  \Bigg[ \rho(a) + \frac{m^4 \mpl^4}{ \mpf^2} \left( \frac{1}{H^2} \right) \Bigg] \, ,
\ee
and the stability constraint is exactly
\be
\frac{m^2 \beta_1}{2} \left( \frac{1}{\mpl^2} + \frac{36 \mpf^4}{\mpf^2 m^4 \beta_1^2} H^4 \right) \ge \frac{12 \mpf^2 H^4}{m^2 \beta_1} \, ,
\ee
which is easily seen to be always satisfied since it simplifies to the trivially satisfied relation
\be
 \left( \frac{1}{\mpl^2} + \frac{12 \mpf^2}{m^4 \beta_1^2} H^4 \right) \ge 0 \, .
\ee
Thus we automatically generate a ghost-free late-time self-accelerating cosmology whose Friedmann equation may be put in the form
\be
H^2 = \frac{1}{6 \mpl^2} \left( \rho(a) + \sqrt{\rho(a)^2 + \frac{12 m^4 \mpl^6}{\mpf^2}} \right) \, , 
\ee
and asymptotes to de Sitter with 
\be
H_{\infty} =  \frac{1}{3^{1/4}}m  {\sqrt{\frac{\mpl}{\mpf}}} \, .
\ee
This represents one subset of the parameter space for which the bound is satisfied at all times. {This specific solution, as well as similar ones with $\beta_3,\beta_3\not=0$, have been studied in \cite{Akrami:2013pna,Akrami:2013ffa}. There the authors employ statistical methods to asses
the viability of bigravity solutions as compared to that of the standard $\Lambda CDM$ cosmology. Quite interestingly, the $\beta_1 \not=0=\beta_{2,3}$ solution discussed, which we have shown to satisfy the stability bound, is on par with $\Lambda CDM$ in their analysis.}

\section{$\Lambda_3$ Decoupling Limit of Bigravity}

\label{Dec}

We shall now derive the complete $\Lambda_3$ decoupling limit for bigravity including vector degrees of freedom. This decoupling limit will allow us to give an independent derivation of the generalized Higuchi bound in bigravity, and also allow us to determine the stability of the vectors. Just as in the massive gravity $\Lambda_3$ decoupling limit, the bigravity decoupling limit focuses on the interactions at the lowest energy scale, those of the helicity-zero mode. 

\subsection{Vierbein formulation}

Our derivation will utilize the vierbein formalism \cite{Hinterbichler:2012cn,Chamseddine1} (see also \cite{Chamseddine2,Nibbelink:2006sz}) and will follow closely the notation of \cite{Ondo:2013wka}. The starting point is to introduce \stu fields for the broken Lorentz invariance and diffeomorphism (diff) invariance. Just as adding a mass term to gravity breaks diffeomorphism invariance, adding a mass term to bigravity breaks two independent copies of the diffeomorphism group down to the single diagonal subgroup. In the Einstein-Cartan (vierbein) formalism, it similarly breaks two copies of local Lorentz transformations (LLTs) down to a single copy. It is helpful to reintroduce these symmetries, and this can be easily achieved using the \stu trick. 

Beginning in unitary gauge, the action for bigravity
\be
S_{\rm bigravity} = \int \d^4 x \, \frac{1}{2}\left[\, \mpl^2 \sqrt{-g} \, R[g]+\mpf^2 \sqrt{-f} \, R[f]-{ m^2}\sum_{n=0}^4 \beta_n \, {\mathcal{U}}_n(X) \right]+ \mathcal{L}_M\, , \label{nomatter}
\end{equation}
can be expressed in vierbein form as
\bea
S_{\rm bigravity} &=& \frac{\mpl^2}{2} \ep_{abcd} \int  \frac{1}{2}E^a \wedge E^b \wedge R^{cd}[E]+ \frac{\mpf^2}{2} \ep_{abcd} \int  \frac{1}{2}F^a \wedge F^b \wedge R^{cd}[f]  \nn  \\
&-& \frac{m^2}{2} \ep_{abcd} \int  \left[ \frac{\b_0}{4!} E^a \wedge E^b \wedge E^c \wedge E^d  + \frac{\b_1}{3!} F^a \wedge E^b \wedge E^c \wedge E^d  \right] \nonumber \\
&-& \frac{m^2}{2} \ep_{abcd} \int \left[\frac{\b_2}{2!2!} F^a \wedge F^b \wedge E^c \wedge E^d  + \frac{\b_3}{3!} F^a \wedge F^b \wedge F^c \wedge E^d \right]  \, .
\eea
Here $F^a$ is the $f$ metric vierbein $f_{\mu\nu} = F_{\mu}^a  F_{\nu}^a \eta_{ab}$ and $E^a$ is the $g$ metric vierbein $g_{\mu\nu} = E_{\mu}^a  E_{\nu}^a \eta_{ab}$. 
We now introduce \stu fields $\Phi^a$ for diffs and $\Lambda^a{}_b$ for LLTs, so that $\Lambda \eta \Lambda^T = \eta$ (see \cite{Ondo:2013wka} for more details) and the bigravity action becomes
\bea
S_{\rm bigravity} &=& \frac{\mpl^2}{2} \ep_{abcd} \int  \frac{1}{2}E^a \wedge E^b \wedge R^{cd}[E]+ \frac{\mpf^2}{2} \ep_{abcd} \int  \frac{1}{2}F^a \wedge F^b \wedge R^{cd}[f]  \nn  \\
&-& \frac{m^2}{2} \ep_{abcd} \int  \left[ \frac{\b_0}{4!} E^a \wedge E^b \wedge E^c \wedge E^d  + \frac{\b_1}{3!} \tilde F^a \wedge E^b \wedge E^c \wedge E^d  \right] \nonumber \\
&-& \frac{ m^2}{2} \ep_{abcd} \int \left[\frac{\b_2}{2!2!} \tilde F^a \wedge \tilde F^b \wedge E^c \wedge E^d  + \frac{\b_3}{3!} \tilde F^a \wedge \tilde F^b \wedge \tilde F^c \wedge E^d \right]  \, ,
\eea
where 
\be
\tilde F^a_{\mu} = \Lambda^a{}_b F^a_{A}(\Phi) \, \partial_{\mu} \Phi^A\, .
\ee
We have apparently broken the symmetry between the two vierbeins/metrics in introducing \stu fields explicitly in one of the vierbeins only. However, since the bigravity action preserves a diagonal copy of diffs and LLTs, we may use this freedom to switch this dependence between the two vierbeins/metrics. 

Following \cite{Ondo:2013wka} we now define
\bea
E^a_{\mu} &=& \de^a_{\mu} + \frac{1}{2 \mpl } h^a_{\mu} \, , \quad \quad F^a_{\mu} = \de^a_{\mu} + \frac{1}{2 \mpf } v^a_{\mu} \nonumber \\
\Lambda^a{}_{b} &=& e^{ \hat{\om}^{a}{}_{ b}} = \delta^a{}_b + \hat{\om}^{a}{}_{ b} + \frac{1}{2} \hat{\om}^{a}{}_{c} \hat{\om}^{c}{}_{b} + \cdots   \nn \\
\hat{\om}^{a}_{\hspace{.2cm} b} &=& \frac{\om^{a}_{\hspace{.2cm} b}}{m \mpl} \nn \\
\p_{\mu}\Phi^a &=& \p_{\mu} \( x^a + \frac{B^a}{m \mpl} + \frac{ \p^{a}\pi}{ \Lambda^3_3} \) 
\eea
and perform the scaling or decoupling limit,
\be
\mpl   \rightarrow  \infty \,, \quad  \mpf   \rightarrow  \infty \, ,\quad m \to  0 
\ee 
while keeping 
\be
\Lambda_3 = (m^2\mpl)^{\frac{1}{3}} \to \text{constant} \quad   \text{and}  \, \quad \, \mpl/\mpf \text{ constant}\, .
\ee
In addition, the scaling is done such that the $\hat \beta_n$ are kept constant where $\beta_n = \mpl^2 \hat \beta_n$. 
The resulting action in the $\Lambda_3$ decoupling limit can be split into two contributions
\be
\lim_{\mpl \rightarrow \infty\, , \, \La_3 \text{ constant}} S_{\rm bigravity} = S_{\rm helicity-2/0}+ S_{\rm helicity-1/0} \,
\ee
where $ S_{\rm helicity-1/0} $ contains only interactions between the helicity-1 and helicity-0 degrees of freedom \cite{Ondo:2013wka}\footnote{As in  \cite{Ondo:2013wka} we take the standard definitions of the Kronecker deltas: $ \de^{\mu \nu \rho \sigma}_{abcd} = \ep^{\mu \nu \rho \sigma}\ep_{abcd} $.  More generally we have $\de^{\mu \nu \rho}_{abc} = \frac{1}{1!} \ep^{\mu \nu \rho d}\ep_{abcd}$ and $\de^{\mu \nu}_{ab} = \frac{1}{2!} \ep^{\mu \nu c d}\ep_{abcd}$.}:
\bea
S_{\rm helicity-1/0}  &=& - \frac{\hat \beta_1}{4} \de^{\mu\nu\rho\si}_{abcd}   \( \frac{1}{2} G_{\mu}^a \om^b{}_{\nu} \de^c_{\rho} \de^d_{\si} + (\de + \Pi)^a_{\mu}
 [ \de^b_{\nu} \om^c{}_{\rho} \om^d{}_{\si}
  +\frac{1}{2}\de^b_{\nu} {\de}^c_{\rho}  \om^d{}_{\al} \om^{\al}{}_{\si} ]
 \) \nonumber \\
&-& \frac{\hat \beta_2}{8} \de^{\mu\nu\rho\si}_{abcd}  \left(  2 G_{\mu}^a (\de + \Pi)^b_{\nu} \om^c {}_{\rho}  \de^d_{\si}+  (\de + \Pi)^a_{\mu} (\de + \Pi)^b_{\nu} [  \om^c{}_{\rho}\om^d{}_{\si}+ \de^d_{\si}\om^c{}_{\al} \om^{\al}{}_{\rho}]   \right)\nonumber \\
&-&  \frac{\hat \beta_3}{24} \de^{\mu\nu\rho\si}_{abcd}  \left( (\de + \Pi)^a_{\mu} (\de + \Pi )^b_{\nu} (\de + \Pi)^c_{\rho} \om^d{}_{\al} \om^{\al}{}_{\si} +3 \om^a {}_{\mu} G_{\nu}^b (\de + \Pi)^c_{\rho} (\de + \Pi )^d_{\si}  \right) \nonumber \, ,
\eea
where
\bea
\om_{ab}  &=&\int^{\infty}_{0} \d u \hspace{.2cm} e^{-2u} e^{ -u\Pi_a{}^{a'} } G_{a'b'} e^{ -u\Pi^{b'}_{\hspace{.2cm}b}} \\
&=&\sum_{n,m} \,  \frac{(n+m)!}{2^{1+n+m}n!m!} (-1)^{n+m}\, \( \Pi^n G \Pi^m \, \)_{ab}\,, \nonumber 
\eea
is the solution of 
\be
G_{ab} = \partial_a B_b - \partial_b B_a =\om_{ac}(\delta + \Pi)^c{}_b + (\delta+ \Pi)_a{}^c  \om_{cb}\,,
\ee
$\Pi_{ab}$ is defined as
\be
 \Pi_{ab} = \frac{\p_a \p_b \pi}{\La^3_3}\,.
\ee
Similarly $S_{\rm helicity-2/0}$ contains the interactions of the helicity-2 and helicity-0 modes and is given by
\bea
S_{\rm helicity-2/0}&=& \int \d^4x \left[  -\frac{1}{4} h^{\mu \nu} \hat{\mathcal{E}}^{\al \b}_{\mu\nu} h_{\al\b} -\frac{1}{4} v^{\mu \nu} \hat{\mathcal{E}}^{\al \b}_{\mu\nu} v_{\al\b} \right.  \nn \\
 &+& \left. \frac{\La_3^3}{2}h^{\mu\nu}(x) X^{\mu\nu} + \frac{\mpl \La_3^3}{2 \mpf} v_{\mu A}[x^a + \La_3^{-3}\partial^a \pi](\eta^A_{\nu} + \Pi^A_{\nu})  Y^{\mu \nu} \right] \, ,
\eea
where $\hat{\mathcal{E}}^{\al \b}_{\mu\nu}$ is the Lichnerowicz operator defined on a background Minkowski space-time with the convention $(\hat{\mathcal{E}}h)_{\mu\nu} =-\frac{1}{2} \left( \Box h_{\mu\nu} - \partial_{\alpha} \partial_{\mu} h^{\alpha}_{\nu} - \partial_{\alpha} \partial_{\nu} h^{\alpha}_{\mu}+ \partial_{\mu} \partial_{\nu} h - \eta_{\mu \nu} (\Box h - \partial_{a} \partial_b h^{ab}) \right)$.
The tensors $X^{\mu\nu}$ and $Y^{\mu\nu}$ are defined by
\be
X^{\mu \nu} = - \frac{1}{2} \sum_{n=0}^4  \frac{\hat \beta_n}{(3-n)! n!}  \ep^{\mu \dots } \ep^{\nu \dots}  (\eta+\Pi)^n \eta^{3-n} \, ,
\ee
and
\be
Y^{\mu \nu} = - \frac{1}{2} \sum_{n=0}^4  \frac{\hat \beta_n}{(4-n)! (n-1)!}  \ep^{\mu \dots } \ep^{\nu \dots}  (\eta+\Pi)^{(n-1)} \eta^{4-n} \, ,
\ee
where we have used a short hand notation in which the indices of $(\eta+\Pi)$ and $\eta$ are contracted between the pairs of Levi-Civita symbols $\ep$ in order. 

In this representation the dependence of the action on $v_{\mu\nu}$ is nontrivial due to $\pi$ dependence in $ v_{\mu A}[x^a + \La_3^{-3}\partial^a \pi](\eta^A_{\nu} + \Pi^A_{\nu}) $ term. We can however undo this with a coordinate transformation in the last term to write an equivalent representation:
\bea
\label{rep}
S_{\rm helicity-2/0}&=& \int \d^4x \left[  -\frac{1}{4} h^{\mu \nu} \hat{\mathcal{E}}^{\al \b}_{\mu\nu} h_{\al\b} -\frac{1}{4} v^{\mu \nu} \hat{\mathcal{E}}^{\al \b}_{\mu\nu} v_{\al\b} \right.  \nn \\
 &+& \left. \frac{\La_3^3}{2}h^{\mu\nu}(x) X^{\mu\nu} + \frac{\mpl \La_3^3}{2 \mpf} v_{\mu \nu}(x^a)\tilde Y^{\mu \nu} \right] \, ,
\eea
where
\be
\tilde Y^{\mu \nu} = - \frac{1}{2} \sum_{n=0}^4  \frac{\hat \beta_n}{(4-n)! (n-1)!}  \ep^{\mu \dots } \ep^{\nu \dots}  \eta^{(n-1)} (\partial Z)^{4-n} \, ,
\ee
and where $(\partial Z)^a_{\nu} = \partial_{\mu}Z^a(x)$ and the function $Z^a(x)$ is defined via the implicit relation
\be
\label{Zdef}
Z^a(x^b + \La_3^{-3}\partial^b \pi(x))=x^a \, .
\ee 
The fact that we have performed the coordinate transformation in only the last term might seem strange, however it is allowed because the integration variable is a dummy variable. Essentially we are using the four dimensional version of the identities
\bea
\int_{-\infty}^{\infty} \d x \left( f(x) + h(x)\right) &=& \int_{-\infty}^{\infty} \d x f(x)+ \int_{-\infty}^{\infty} \d x h(x)= \int_{-\infty}^{\infty} \d x f(x)+\int_{-\infty}^{\infty} \d Z \,  h(Z)  \nn  \\
& =& \int_{-\infty}^{\infty} \d x \left( f(x)+\frac{dZ(x)}{dx} h(Z(x)) \right) \, .
\eea
where $Z(x)$ is a monotonic function satisfying $Z(\pm \infty)=\pm \infty$. 

To elucidate the meaning of this remember that the diff \stu fields are defined in the decoupling limit as $\Phi^a(x) = x^a + \La_3^{-3}\partial^b \pi(x)$, thus the relation (\ref{Zdef}) is
\be
Z^a(\Phi^b(x)) = x^a \, ,
\ee
in other words the function $Z^a$ is the inverse function, i.e. inverse coordinate transformation to $\Phi^a$. The function $\Phi^a$ provides a map from the coordinates of the metric $g$ to the metric $f$, so the function $Z^a$ gives the inverse map.  \\

An important property of this function is that it satisfies $\partial_a Z_b = \partial_b Z_a$. To prove this, differentiate equation (\ref{Zdef}), to give\footnote{To be clear on notation since we will utilize this trick several times, $[\partial_a A](f^b(x))$ indicates the partial derivative evaluated at the point $f^b(x)$ to be distinguished from $\partial_a (A(f(x))  = (\partial_a f^b(x)) [\partial_b A](f(x))$ which is the partial derivative with respect to $x^a$ of the function evaluated at $f^b(x)$.}
\be
\label{SW}
[\partial^c Z^a](x+\La_3^{-3}\partial \pi(x) ) \left[\eta_{bc}+ \La_3^{-3} \partial_{b}\partial_c \pi \right] = \eta^a_{b} \, .
\ee
Inverting gives, in matrix notation,
\be
\label{inverting}
[\partial Z](x+\La_3^{-3}\partial \pi(x) ) = [\eta+ \Pi]^{-1} \, 
\ee
and, since $\Pi^T=\Pi$, we have $[\partial Z]^T= [\partial Z]$ which is the desired relation.

Given this relation, we have the identity that
\be
\partial_{\mu} \tilde Y^{\mu\nu}=0 \, . 
\ee
which is necessary to preserve linear diff invariance $v_{\mu\nu} \rightarrow v_{\mu\nu} + \partial_{\mu} \xi_{\nu}+\partial_{\nu} \xi_{\nu}$.
This follows because of the antisymmetry properties of the $\ep$ tensors, every time two derivatives act on $Z$, the antisymmetry forces this term to zero since we can use the $\partial_a Z_b - \partial_b Z_a=0$ property to ensure that the two derivatives are contracted with the same $\ep$ tensor.

Since $Z^a$ contains only one scalar degree of freedom $\pi(x)$, and since $\partial_a Z_b - \partial_b Z_a=0$, it must be possible to express it as
\be
Z^a(x) = x^a + \frac{1}{\La_3^3} \partial^a \rho(x) \, .
\ee
In other words, associated with $\pi(x)$ is a conjugate or dual variable $\rho(x)$ which is related by a nonlocal field redefinition. We can determine this function by iterating the equation (\ref{Zdef}).
To order $1/\Lambda_3^6$ the map is given by
\be
\rho(x) = -\pi(x) + \frac{1}{2 \La_3^3} (\partial_b \pi(x))^2- \frac{1}{2 \La_3^6} \partial^a \pi(x) \partial^b \pi(x) \partial_a \partial_b \pi(x) + \dots \, .
\ee
This redefinition is nonlocal because there are an infinite number of terms. This non locality is actually crucial to evading the Ostrogradski theorem\footnote{A field redefinition that includes a finite number of higher time derivatives will always introduce higher time derivatives into the action and hence a ghost by virtue of the Ostrogradski theorem (see \cite{Chen:2012au} for a recent discussion). The only way to evade the ghost is if there are an infinite number of terms, i.e. if the transformation is nonlocal. In this case the Ostrogradski analysis does not apply. If it can be shown that the equations of motion are second order after the field redefinition, and if the field redefinition is invertible, then there cannot be a ghost present. In the present case the absence of any Ostrogradski type ghost is an automatic consequence of its derivation from the ghost-free bigravity models. We shall make this explicit in \cite{duality}.}. Thus we should think of $\rho(x)$ as the helicity-0 mode defined from the perspective of the $f$ metric just as $\pi(x)$ is the helicity-0 mode from the perspective of the $g$ metric. We may use either variable as the fundamental degree of freedom, but the map between the two, which is the map between the two metrics, is nonlocal. One may worry that the resulting equations of motion are nonlocal, but this is not the case \cite{duality}.

The interactions of the second metric $f$ to the helicity-zero mode are most naturally expressed in terms of identical interactions to $g$ but with $\pi(x)$ replaced by $\rho(x)$. 
\be
\tilde Y^{\mu \nu} = - \frac{1}{2} \sum_{n=0}^4  \frac{\hat \beta_n}{(4-n)! (n-1)!}  \ep^{\mu \dots } \ep^{\nu \dots}  \eta^{(n-1)} \left(\eta +\Sigma \right)^{4-n} \, .
\ee
where $\Sigma_{\mu\nu} =\partial_{\mu} \partial_{\nu} \rho/\La_3^3$. 

\subsection{Dual Vainshtein effect}

In this representation (\ref{rep}) we see that the $\Lambda_3$ decoupling limit is invariant under the interchange of $\pi \leftrightarrow \rho$,  $\Pi \leftrightarrow \Sigma$, $\beta_n \leftrightarrow \beta_{4-n}$, $v_{\mu\nu} \leftrightarrow h_{\mu\nu}$, $\mpl \leftrightarrow \mpf$. This duality is reminiscent of electric-magnetic (Heaviside) duality.
Under this interchange gravitational charges coupled to $f$ are replaced by gravitational charges coupled to $g$ and vice versa, analogous to the interchange of charges and monopoles. Electric-magnetic duality is also a nonlocal map when written in terms of the gauge potentials despite the fact that it maps two local theories \cite{Deser:1976iy}. There is also an interesting variant on the idea of strong/weak coupling duality $g \leftrightarrow 1/g$ \cite{Montonen:1977sn}. To see this we can write equation (\ref{SW}) in the form
\be
[\eta+ \Pi(x)] = \frac{1}{[\eta+\Sigma(\Phi)]} \, \quad \text{or} \quad [\eta+\Sigma(\Phi)] = \frac{1}{[\eta+\Pi(x)]} 
\ee
Deep inside the usual Vainshtein region we have $\Pi(x) \gg 1$. In the normal Vainshtein effect the kinetic term for the fluctuations of $\pi$ are then renormalized in such a way that the fluctuations become weakly coupled to sources, which is how the Vainshtein mechanism \cite{Vain} resolves the vDVZ discontinuity \cite{vDVZ}. In terms of the dual Galileon field this maps onto a region for which $\Sigma \sim -\eta$. This is at the border of the weak/strong coupling region in terms of the dual Galileon for sources minimally coupled to $\rho$.  Thus while this is not exactly a weak/strong coupling duality, it does relate two apparently very different regions of the theory.  We shall discuss this duality in more detail elsewhere \cite{duality}.

This last point brings us to one of the main physical conclusions of this work. We have seen that resolving the Higuchi-Vainshtein tension in bigravity has forced us to consider the regime $H/H_f = b/a \ll \mpl/\mpf $, a region that does not exist in massive gravity. Since we usually have in mind that $\mpf > \mpl$ this implies $b/a \ll 1$.
As we shall see in {\it Section}~\ref{BiDec}, the profile of the Galileon for cosmological solutions is of the form 
\be
\pi =\frac{\La_3^3}{2} \left( \frac{b}{a}-1 \right) x^{\mu} x_{\mu}
\ee
which is equivalent to 
\be
[1+ \Pi ]_{\mu \nu} = \frac{b}{a} \eta_{\mu \nu} \, .
\ee
Thus in the region $H/H_f = b/a \ll 1$ where we recover continuity with GR in the Friedmann equation, we have precisely $\Pi \sim -\eta$. According to the above duality this corresponds to $\Sigma \gg \eta$. Thus it is the dual Galileon $\rho$ and not the usual Galileon $\pi$ which exhibits the large kinetic term (wavefunction) renormalization characteristic of the Vainshtein effect. We can thus call the region where $\Sigma \gg \eta $ the dual Vainshtein region which is conjugate to usual Vainshtein region $\Pi \gg \eta$ \cite{duality}. 

Whether we are in the Vainshtein region or the dual Vainshtein region depends essentially on which of the two metrics dominates the mass eigenstates. In the usual massive gravity limit, it is always the $g$ metric which is massive, and so the Vainshtein effect is always dominated by $\pi$. However, in the bigravity case, around a dynamical background, the mass eigenstates can change with time. In the present case, even though $\mpf \gg \mpl$, the fact that $H \mpf \ll H_f \mpl$ is enough to ensure that it is the $f$ metric and not the $g$ metric which is mostly build out of the massive graviton eigenstate. This can be seen from the {\it Appendix}~\ref{Appendix} where we have $F_3 \gg F_1$ in this region. It is for this reason that it is the dual field $\rho$ which governs the Vainshtein effect here. The Higuchi-Vainshtein conflict is resolved because at early times, matter which is minimally coupled to $g$ (which has been our assumption throughout), is mostly coupled to the massless eigenstate, and hence we recover a GR-like Friedmann equation at early times. It is only at late-times, when the `dark energy' component of the Friedmann equation comes to dominate, that the matter begins to couple equally to the massless and the massive graviton. This means that in resolving the Higuchi-Vainshtein tension, we have made it that for most of the history of the universe the force of gravity was propagated principally by the massless graviton. Only at late times does the massive graviton take over. It is thus no surprise that this resolution could not be seen using massive gravity alone. 

\subsection{Decoupling Limit of Massive Gravity for Arbitrary Reference Metrics}

The bigravity $\La_3$ decoupling limit can also be used as a tool to derive the massive gravity decoupling limit for non-Minkowski reference metrics. The only explicit cases that have been studied are in (anti-) de Sitter \cite{deRham:2012kf}, (see also \cite{Mirbabayi:2011aa} related discussion). The results of  \cite{deRham:2012kf} show that accounting for a de Sitter reference metric simply adds new Galileon type interaction for the helicity-0 mode in the decoupling limit. This is consistent with the general result which we now derive. The only place the $f$ metric comes in is in the helicity-2/helicity-0 interactions
\bea
S_{\rm helicity-2/0}&=& \int \d^4x \left[  -\frac{1}{4} h^{\mu \nu} \hat{\mathcal{E}}^{\al \b}_{\mu\nu} h_{\al\b} -\frac{1}{4} v^{\mu \nu} \hat{\mathcal{E}}^{\al \b}_{\mu\nu} v_{\al\b} \right.  \nn \\
 &+& \left. \frac{\La_3^3}{2}h_{\mu\nu}(x) X^{\mu\nu} + \frac{\mpl \La_3^3}{2 \mpf} v_{\mu A}[x^a + \La_3^{-3}\partial^a \pi](\eta^A_{\nu} + \Pi^A_{\nu})  Y^{\mu \nu} \right] \, .
\eea
To obtain the massive gravity decoupling limit, one can add an external source for $v_{\mu\nu}$ that generates a background for $\bar v_{\mu\nu}$ and then take the limit $\mpf \rightarrow \infty$, keeping $U_{\mu\nu}(x)=\mpl\, \bar v_{\mu\nu}(x)/\mpf$ fixed. In the limit $\mpf/\mpl \rightarrow \infty$, the fluctuations of $v_{\mu\nu}$ decouple and so we obtain the $\Lambda_3$ decoupling limit
\bea
S_{\rm helicity-2/0}&=& \int \d^4x \left[  -\frac{1}{4} h^{\mu \nu} \hat{\mathcal{E}}^{\al \b}_{\mu\nu} h_{\al\b} \right.  \nn \\
 &+& \left. \frac{\La_3^3}{2}h^{\mu\nu}(x) X^{\mu\nu} + \frac{\La_3^3}{2} U_{\mu A}[x^a + \La_3^{-3}\partial^a \pi](\eta^A_{\nu} + \Pi^A_{\nu})  Y^{\mu \nu} \right] \, ,
\eea
for massive gravity on a reference metric which is expressed in a locally inertial frame as $f_{\mu\nu}(x) = \eta_{\mu\nu} + U_{\mu\nu}(x)/\mpl$. It might seem a restriction that the decoupling limit is only valid for reference metrics which are perturbatively close to Minkowski. However, this is a necessary condition since we must scale the curvature as $R[f] \sim m^2 \sim 1/\mpl$ in taking the limit. In other words, the curvature radius tends to infinity in the limit. We can still use this to describe a portion of any space-time by going into a locally inertial frame for the $f$ metric, where locally, well inside the curvature radius, any space-time may be well described by Minkowski plus small perturbations. 

To give a concrete example, consider the case where the reference metric is de Sitter. De Sitter spacetime with Hubble constant $H_f$ can be expressed in a locally inertial frame as
\be
f_{\mu\nu}(x) = \eta_{\mu\nu}   H_f^2 \, x_{\mu}x_{\nu} + {\cal O}(x^4) \, .
\ee
In the decoupling limit we send $H_f \sim m \rightarrow 0$ so that the Hubble horizon tends to infinity. Nevertheless, $H_f$ still makes nonzero corrections to the decoupling limit Lagrangian due to the fact that we are simultaneously sending $\mpl \rightarrow \infty$. 

Explicitly, these are
\bea
S^{\text{de Sitter}}_{\rm helicity-2/0}&=& \int \d^4x \left[  -\frac{1}{4} h^{\mu \nu} \hat{\mathcal{E}}^{\al \b}_{\mu\nu} h_{\al\b} + \frac{\La_3^3}{2}h^{\mu\nu}(x) X^{\mu\nu}  \right.  \nn \\
 &+& \left.  \frac{\La_3^3}{2} \mpl H_f^2 \, (x_{\mu} + \La_3^{-3}\partial_{\mu} \pi)    (x_{A} + \La_3^{-3}\partial_{A} \pi)  (\eta^A_{\nu} + \Pi^A_{\nu})  Y^{\mu \nu} \right] \, .
\eea
Although not immediately apparent, one can show that the new $H_f^2$ terms are all equivalent to Galileon interactions, as derived by a different method in \cite{deRham:2012kf}. To see this, define $\hat \pi = \pi/\La_3^3 + x^2/2$ so that $\partial_{\mu} \hat \pi = x_{\mu} +  \La_3^{-3}\partial_{\mu} \pi$ and $\partial \partial \hat \pi = \eta + \Pi$. Then we have:
\be
S^{\text{de Sitter}}_{\rm helicity-2/0}= \int \d^4x \left[  -\frac{1}{4} h^{\mu \nu} \hat{\mathcal{E}}^{\al \b}_{\mu\nu} h_{\al\b} + \frac{\La_3^3}{2}h^{\mu\nu}(x) X^{\mu\nu}  +   \frac{\La_3^3}{2} \mpl H_f^2 \,   \partial_{\mu} \hat \pi \partial_{A} \hat \pi  \Pi^A_{\nu} Y^{\mu \nu} \right] \, .
\ee
Integrating by parts using $ \partial_{A} \hat \pi  \Pi^A_{\nu}  = \frac{1}{2} \partial_{\nu} ( (\partial_{A} \hat \pi)^2)$ and using the fact that $Y^{\mu \nu}$ is conserved, we have:
\be
S^{\text{de Sitter}}_{\rm helicity-2/0}= \int \d^4x \left[  -\frac{1}{4} h^{\mu \nu} \hat{\mathcal{E}}^{\al \b}_{\mu\nu} h_{\al\b} + \frac{\La_3^3}{2}h^{\mu\nu}(x) X^{\mu\nu}  -  \frac{\La_3^3}{4} \mpl H_f^2 \,    (\partial_{A} \hat \pi)^2  \Pi_{\mu \nu} Y^{\mu \nu} \right] \, ,
\ee
which, given the definition of $Y^{\mu\nu}$, is the same as:
\bea
S^{\text{de Sitter}}_{\rm helicity-2/0}&=& \int \d^4x \left[  -\frac{1}{4} h^{\mu \nu} \hat{\mathcal{E}}^{\al \b}_{\mu\nu} h_{\al\b} + \frac{\La_3^3}{2}h^{\mu\nu}(x) X^{\mu\nu}  \right.  \\
 &+& \left.   \frac{\La_3^3}{8}  \mpl H_f^2 \,  \sum_{n=0}^4  \frac{\hat \beta_n}{(4-n)! n!}  (\partial_{A} \hat \pi)^2   \ep \ep \left( (\eta + \Pi)^n \eta^{4-n}  -(\eta+\Pi)^{n-1}\eta^{5-n}  \right)\right] \, . \nn
\eea
These extra terms are manifestly Galileon interactions build out of $\hat \pi$ since $\eta + \Pi = \partial \partial \hat \pi$. However, as is well known, performing the shift $\hat \pi = \pi/\La_3^3 + x^2/2$ and integrating by parts just modifies the coefficients of the different Galileon operators and, on so doing, we obtain the representation of   \cite{deRham:2012kf}. Finally we note that the anti-de Sitter case is covered by replacing $H_f^2$ with $-1/l^2$ where $l$ is the AdS curvature length scale.

\section{Bigravity Decoupling Limit Derivation of Stability Bound}

\label{BiDec}

We will now use the bigravity $\Lambda_3$ decoupling limit to give an independent derivation of the generalized Higuchi bound. For simplicity, let us begin by concentrating on the case when both metrics are de Sitter. As we have seen, since the precise form of matter and $\dot H $ drop out of the bound, it is sufficient to look at this case. This case is of course easy to perform exactly as we do in the {\it Appendix}~\ref{Appendix}, however we follow through the decoupling limit steps here since it better elucidates the physics of the Vainshtein mechanism, and how to go beyond de Sitter and linear perturbations. We will explain below how this is generalized to the case where both metrics are FRW. This analysis will also enable us to look at the potential existence of other branches of solutions and the vector perturbations. 

\subsection{de Sitter on de Sitter}

To begin with let us focus on the case where both metrics are de Sitter. In the $\Lambda_3$ decoupling limit this is described by background solutions for the metric perturbations of the form
\be
\bar h_{\mu\nu} = {\mpl} H^2 \, x_{\mu}x_{\nu} \, ,
\ee
and
\be
\bar v_{\mu\nu} = {\mpf}H_f^2 \, x_{\mu}x_{\nu} \, .
\ee
The related background solution for $\pi$ written in a way to make easy contact with our previous results
\be
\bar \pi  = \frac{\La_3^3}{2} x^{\mu}x_{\mu} \left( \frac{b}{a}-1\right) ,
\ee
so that 
\be
\eta+ \bar \Pi = \frac{b}{a} \eta .
\ee
The related background for the vector fields is zero $\bar B_{\mu}=0$. 

In this section we will work in the representation 
\bea
S_{\rm helicity-2/0}&=& \int \d^4x \left[  -\frac{1}{4} h^{\mu \nu} \hat{\mathcal{E}}^{\al \b}_{\mu\nu} h_{\al\b} -\frac{1}{4} v^{\mu \nu} \hat{\mathcal{E}}^{\al \b}_{\mu\nu} v_{\al\b} \right.  \nn \\
 &+& \left. \frac{\La_3^3}{2}h^{\mu\nu}(x) X^{\mu\nu} + \frac{\mpl \La_3^3}{2 \mpf} v_{\mu \nu}(x)  \tilde Y^{\mu \nu} \right] \, ,
\eea
for which the helicity-0 interactions are simple, and the only nontrivial part of the action is the term with $v_{\mu A}[x^a + \La_3^{-3}\partial^a \pi]$.

The equations of motion that follow from this action are, for the helicity-2 modes,
\be
\hat{\mathcal{E}}^{\al \b}_{\mu\nu} h_{\al\b} =  {\La_3^3} X_{ab} \, .
\ee
\be
\hat{\mathcal{E}}^{\al \b}_{\mu\nu} v_{\al\b} = {\La_3^3} \tilde Y_{ab} \, .
\ee
For the helicity-0 mode it requires a little more work. We will vary both $\pi$ and $\rho$ and use the basic connection $\delta \rho(x+ \partial \pi/\La_3^3) = - \delta \pi(x)$ to obtain a single equation. 
To prove this relation, starting with $Z^a(\Phi^b)=x^a$ we have on varying
\be
[\partial^a \delta \rho](\Phi) + [\partial_b Z^a] \partial_b \delta \pi(x) =0 \, . 
\ee
which can be rewritten as 
\be
[\eta+\Pi] ^a_b[\partial^b \delta \rho](\Phi) + \partial^a \delta \pi(x) =0 \, . 
\ee
using Eq.~(\ref{inverting}). This is equivalent to
\be
\partial^a \left( \delta \rho(\Phi) + \delta \pi(x) \right) =0 \, ,
\ee
which gives us our desired result. 

The formal steps in the derivation of the equations of motion for the helicity-0 mode are as follows: We vary the action as
\be
\delta S_{\rm helicity-2/0} = \int \d^4 x \left[  \frac{\La_3^3}{2} h^{\mu\nu}(x) \frac{\delta X^{\mu\nu}}{\delta \pi} (x) \delta \pi(x) +  \frac{\mpl \La_3^3}{2 \mpf} v^{\mu\nu}(x) \frac{\delta \tilde Y^{\mu\nu}}{\delta \rho} (x) \delta \rho(x) \right] \, .
\ee
We then change the dummy integration variables in the second part of the integral from $x \rightarrow x + \partial \pi/\La_3^3$, which gives a Jacobian factor of $|{\rm Det}[\eta+ \Pi]|$ so that
\bea
\delta S_{\rm helicity-2/0} &=& \int \d^4 x \left[  \frac{\La_3^3}{2} h^{\mu\nu}(x) \frac{\delta X^{\mu\nu}}{\delta \pi} (x) \delta \pi(x) \right. \\
&+& \left. |{\rm Det}[\eta+ \Pi]| \frac{\mpl \La_3^3}{2 \mpf} v^{\mu\nu}( x + \partial \pi/\La_3^3) \frac{\delta \tilde Y^{\mu\nu}}{\delta \rho} ( x + \partial \pi/\La_3^3) \delta \rho( x + \partial \pi/\La_3^3) \right] \, . \nn
\eea
Finally we use the fact that $\delta \rho(x+ \partial \pi/\La_3^3) = - \delta \pi(x)$ to write this as
\bea
\delta S_{\rm helicity-2/0} &=& \int \d^4 x \left[  \frac{\La_3^3}{2} h^{\mu\nu}(x) \frac{\delta X^{\mu\nu}}{\delta \pi} (x) \delta \pi(x) \right. \\
&-&  \left. |{\rm Det}[\eta+ \Pi]| \frac{\mpl \La_3^3}{2 \mpf} v^{\mu\nu}( x + \partial \pi/\La_3^3) \frac{\delta \tilde Y^{\mu\nu}}{\delta \rho} ( \Phi(x)) \delta \pi(x) \right] \, . \nn
\eea
In this form we can then argue that the coefficient of $\delta \pi(x)$ vanishes to give the equation of motion. 

Following through these steps in detail we obtain the equation
\bea
 \sum_{n=1}^3 \frac{\hat \beta_n}{(3-n)! (n-1)!}  \, && \left[  R^{\mu \nu}_{h \, ab}(x)  \ep^{ab..} \ep_{\mu \nu ..}  (\eta+\Pi(x))^{n-1} \eta^{3-n}-  \right. \\
&&  \left.  |{\rm Det}[\eta+ \Pi]| R^{\mu \nu}_{v \, ab} ( x + \partial \pi/\La_3^3)   \ep^{ab..} \ep_{\mu \nu ..} \eta^{n-1} ([\eta+ \Pi(x)]^{-1} )^{3-n}  \right] \nn
\eea
where $R^{\mu \nu}_{h \, ab}(x)$ is the linearized Riemann curvature of $h_{\mu\nu}/\mpl$ and $R^{\mu \nu}_{v \, ab}(x)$ is the same for $v_{\mu\nu}(x)/\mpf$. 

Putting all this information together, and plugging  in the background solution, the equations of motion in the decoupling limit imply the background Friedmann equations
\bea
H^2  = \frac{1}{3 \mpl^2}  \rho + \sum_{n=0}^3 \frac{\hat \beta_n}{(3-n)! n!} \left( \frac{b}{a}\right)^n  \, , \\
H_f^2 = \frac{\mpl^2}{\mpf^2} \sum_{n=0}^3 \frac{\hat \beta_{n+1}}{(3-n)! n!} \left( \frac{b}{a}\right)^{(n-3)} \, ,
\eea
together with the equation of motion for $\bar \pi$ which becomes
\be
\left( \sum_{n=0}^2  \frac{\hat \beta_{n+1}}{(2-n)! n!} \left( \frac{b}{a} \right)^{n+1}   \right) \left( \frac{H^2}{b^2} - \frac{H^2_a}{a^2} \right) =0 \, .
\ee
These are in fact the exact equations, reflecting again the fact that the decoupling limit captures all the relevant features of the full cosmology.

This leads to two branches of solutions, the normal one which we have dealt with in the main text for which we set
\be
\left( \frac{H}{b} - \frac{H_f}{a} \right) =0  \, , \quad \textit{normal branch} \, 
\ee
(branch 2 in the language of \cite{Crisostomi}) and a `alternative branch' which comes from choosing
\be
\label{branch2}
\left( \sum_{n=0}^2  \frac{\hat \beta_{n+1}}{(2-n)! n!} \left( \frac{b}{a} \right)^{n+1}   \right) =0  \, , \quad  \textit{alternative branch} \, 
\ee
(branch 1 in the language of \cite{Crisostomi}).
However, as we shall see shortly, the alternative branch is infinitely strongly coupled due to the fact that the vector kinetic term vanishes (see also \cite{Crisostomi}). This is consistent with the result found in massive gravity \cite{deRham:2010tw}. This means that this branch of solutions should be discarded as it does not lie within the regime of the effective field theory. 

Now, consider perturbations about this background solution expressed in the following manner
\bea
&& h_{\mu\nu}(x)  = \bar h_{\mu\nu}(x) + \delta h_{\mu\nu}(x) \, , \quad \quad  v_{\mu\nu}(x)  = \bar v_{\mu\nu}(x) +  \delta v_{\mu\nu}\(x\)  \, ,\nn \\ 
&&  \pi(x) = \bar \pi(x) +  \delta \pi(x) \, ,\quad \delta \rho(b x/a) = - \delta \pi(x) \, , \quad  B_{\mu}(x) = 0 + \delta B_{\mu}(x).
\eea
Beginning with the action for the helicity-2 mode coupled to the helicity-0 mode expanded to second order, we have 
\bea
S^{(2)}_{\rm helicity-2/0} &=& \int \d^4x \left[  -\frac{1}{4} \delta h^{\mu \nu} \hat{\mathcal{E}}^{\al \b}_{\mu\nu} \delta h_{\al\b} -\frac{1}{4} \delta v^{\mu \nu} \hat{\mathcal{E}}^{\al \b}_{\mu\nu} \delta v_{\al\b} \right.  \nn \\
 &-&  \frac{\La_3^3}{4} \sum_{n=0}^4 \frac{\hat \beta_n }{(3-n)! (n-1)!} \left( \frac{b}{a}\right)^{n-1} \left[  \ep \ep \, \delta h \, \delta \Pi \, \eta \, \eta  \right] \nn \\
 &-& \frac{\La_3^3}{4} \sum_{n=0}^4\frac{\hat \beta_n }{(3-n)! (n-2)!} \left( \frac{b}{a}\right)^{n-2}  \left[  \ep \ep \, ( \partial \partial \bar h) \, \delta \Pi \, \eta \right] \delta \pi  \nn \\
 &-&  \frac{\La_3^3 \mpl}{4 \mpf} \sum_{n=0}^4 \frac{\hat \beta_n }{(3-n)! (n-1)!} \left( \frac{b}{a}\right)^{3-n} \left[  \ep \ep \, \delta v \, \delta \Sigma \, \eta \, \eta  \right] \nn \\
 &-& \left.  \frac{\La_3^3 \mpl }{4 \mpf} \sum_{n=0}^4\frac{\hat \beta_n }{(2-n)! (n-1)!} \left( \frac{b}{a}\right)^{2-n}  \left[  \ep \ep \, ( \partial \partial \bar v) \, \delta \Sigma \, \eta \right] \delta \rho  \right] \, .
 \eea
Here $\beta_0$ and $\beta_4$ are nonzero and are fixed by the tadpole cancellation requirement, i.e. the requirement that the two de Sitter spacetimes are solutions in the absence of matter:
\bea
&&H^2 = \sum_{n=0}^3 \frac{\beta_n}{\mpl^2(3-n)! n!} \left( \frac{H}{H_f} \right)^n  \, , \\
&&H_f^2 = \sum_{n=0}^3 \frac{\beta_{n+1}}{\mpf^2(3-n)! n!} \left( \frac{H}{H_f} \right)^{n-3}  \, .
\eea
We then choose to diagonalize as
\be
\delta h_{\mu \nu}(x)  = -\La_3^3  \sum_{n=0}^4 \frac{\hat \beta_n }{(3-n)! (n-1)!}  \left( \frac{b}{a}\right)^{n-1} \eta_{\mu \nu}  \delta \pi (x) + \delta \hat h_{\mu \nu}(x) \, ,
\ee
and
\be
\delta v_{\mu \nu}(x)  = -\La_3^3  \frac{\mpl}{\mpf} \sum_{n=0}^4 \frac{\hat \beta_n }{(3-n)! (n-1)!}  \left( \frac{b}{a}\right)^{3-n} \eta_{\mu \nu}  \delta \rho (x) + \delta \hat v_{\mu \nu}(x) \, .
\ee
These diagonalizations decouple the helicity-2 mode from the helicity-0 mode, and give an action of the schematic form (after integration by parts)
\be
S^{(2)}_{\rm helicity-2/0}= \int \d^4x \left[  -\frac{1}{4} \delta \hat h^{\mu \nu} \hat{\mathcal{E}}^{\al \b}_{\mu\nu} \delta \hat h_{\al\b} -\frac{1}{4} \delta \hat v^{\mu \nu} \hat{\mathcal{E}}^{\al \b}_{\mu\nu} \delta \hat v_{\al\b} - \frac{\mu_1}{2} (\partial \pi)^2(x) - \frac{\mu_2}{2} (\partial \rho)^2 (x) \right]  \, ,
\ee
where the constants $\mu_i$ depend on the $\beta_n$, $b/a$ and $H$ and $H_f$. At this point we change the dummy variables from $x \rightarrow (b/a) x$ in the final term to give
\be
S^{(2)}_{\rm helicity-2/0}= \int \d^4x \left[  -\frac{1}{4} \delta \hat h^{\mu \nu} \hat{\mathcal{E}}^{\al \b}_{\mu\nu} \delta \hat h_{\al\b} -\frac{1}{4} \delta \hat v^{\mu \nu} \hat{\mathcal{E}}^{\al \b}_{\mu\nu} \delta \hat v_{\al\b} - \frac{\mu_1}{2} (\partial \pi)^2(x) - \frac{b^4 \mu_2}{2a^4} \([\partial \rho]\(\frac{b}{a} x \) \)^2 \right]  \, ,
\ee
which on using $\delta \rho(bx/a) = - \delta \pi(x)$ gives
\be
S^{(2)}_{\rm helicity-2/0}= \int \d^4x \left[  -\frac{1}{4} \delta \hat h^{\mu \nu} \hat{\mathcal{E}}^{\al \b}_{\mu\nu} \delta \hat h_{\al\b} -\frac{1}{4} \delta \hat v^{\mu \nu} \hat{\mathcal{E}}^{\al \b}_{\mu\nu} \delta \hat v_{\al\b} - \frac{1}{2}\( \mu_1+ \mu_2 \( \frac{b}{a} \)^2\)(\partial \pi)^2 \right] \, .
\ee
This finally gives the bound as
\be
 \mu_1+ \mu_2 \( \frac{b}{a} \)^2 \ge 0  \, ,
\ee
which is equivalent to the derived relation
\be
\tilde{m}^2 \left(H^2+ \frac{\mpl^2 H_f^2}{\mpf^2} \right) -2H^4 \geq 0 \, .
\ee

\subsection{Generalization to FRW on FRW} 

Although we will not carry through the full details, this analysis may easily be extended to generic spatially flat FRW background metrics. 
The appropriate ansatz is (this is in a different gauge than the de Sitter case but, since the action is invariant under linear diffs, it will give the same result on de Sitter)
\be
\bar h_{\mu\nu} \d x^{\mu} \d x^{\nu} = -\frac{\mpl}{2} H^2 \vec{x}^2 \d \vec{x}^2  + {\mpl}( \dot H +H^2) \, \vec x^2 \d t^2 \, ,
\ee
and
\be
\bar v_{\mu\nu} \d x^{\mu} \d x^{\nu}  = -\frac{\mpf}{2} H_f^2 \vec{x}^2 \d \vec{x}^2  + {\mpf}( \dot H_f +H_f^2) \, \vec x^2 \d t^2 \,.
\ee
The appropriate ansatz for $\pi$ is 
\be
\pi(\vec x,t) = C(t) + \frac{\La_3^3}{2}  \left( \frac{b(t)}{a(t)}-1 \right) \vec x^2 + {\cal O}(\vec x^4) \, .
\ee
By substituting this ansatz into 
\be
\hat{\mathcal{E}}^{\al \b}_{\mu\nu} h_{\al\b} =  {\La_3^3} X_{ab} \, .
\ee
and using our knowledge of the background Friedmann equations, we will see that $C(t)$ depends on both $b/a$, $\dot{H}$ and $H^2$. 

Fortunately we do not need to know $C(t)$. It is straightforward to see that the $\dot H$ and $\dot H_f$ dependence will not affect the resulting coefficient of the kinetic term for $\pi$. The reason is that due to the $\ep$ structure, only $\bar h_{ij}$, $\bar v_{ij}$ and $\bar \Pi_{ij}$ couple to the terms with two time of derivatives of $\delta \pi$. But, since $\bar h_{ij}$ and $\bar v_{ij}$ do not contain $\dot  H$ and $\dot H_f$, they do not affect the kinetic term. This gives a simple explanation of the result first observed in \cite{Fasiello:2012rw} that the bound on FRW is identical to the bound on de Sitter. 
In fact, it is the same structure that makes massive gravity and bigravity ghost-free that guarantees this result. Of course, the gradient terms are modified by $\dot{H}$ and so an important question to address is whether there exist any gradient or tachyonic instabilities. In \cite{Crisostomi} it was pointed out that there can exist a Jeans like instability in the scalar sector at sub-horizon scales.  In practice this means that the short wavelength scalar fluctuations will become strong coupled and cannot be treated using linearized perturbation theory. It should be possible to use the decoupling limit theory to analyze what happens to them. {Suffice to say that during any quasi-de Sitter period $\dot{H} \ll H^2$ gradient instabilities are not a concern. Notice also that the dS result of \cite{Crisostomi}, although expressed in a different fashion, is compatible with the one obtained here. }

\subsection{Vector modes}

One immediate advantage of using the decoupling limit derivation is that we can also address the question of bounds coming from the stability of the vector and tensor modes. The tensor modes are always manifestly stable as the diagonalized tensor sector is identical to two copies of linearized GR. The vectors on the other hand deserve some attention. Since there are no vectors turned on in the background, the relevant part of the action is 
\bea
S_{\rm helicity-1/0}  &=& - \frac{\hat \beta_1}{4} \de^{\mu\nu\rho\si}_{abcd}   \( \frac{1}{2} G_{\mu}^a \om^b{}_{\nu} \de^c_{\rho} \de^d_{\si} + (\de + \bar \Pi)^a_{\mu}
 [ \de^b_{\nu} \om^c{}_{\rho} \om^d{}_{\si}
  +\frac{1}{2}\de^b_{\nu} {\de}^c_{\rho}  \om^d{}_{\al} \om^{\al}{}_{\si} ]
 \) \nonumber \\
&-& \frac{\hat \beta_2}{8} \de^{\mu\nu\rho\si}_{abcd}  \left(  2 G_{\mu}^a (\de + \bar \Pi)^b_{\nu} \om^c {}_{\rho}  \de^d_{\si}+  (\de + \bar \Pi)^a_{\mu} (\de + \bar \Pi)^b_{\nu} [  \om^c{}_{\rho}\om^d{}_{\si}+ \de^d_{\si}\om^c{}_{\al} \om^{\al}{}_{\rho}]   \right)\nonumber \\
&-&  \frac{\hat \beta_3}{24} \de^{\mu\nu\rho\si}_{abcd}  \left( (\de + \bar  \Pi)^a_{\mu} (\de + \bar \Pi )^b_{\nu} (\de + \bar \Pi)^c_{\rho} \om^d{}_{\al} \om^{\al}{}_{\si} +3 \om^a {}_{\mu} G_{\nu}^b (\de +\bar  \Pi)^c_{\rho} (\de + \bar \Pi )^d_{\si}  \right) \nonumber \, .
\eea
Since the background $\bar \Pi$ is diagonal (to lowest order in $\vec{x}$ in FRW case) we can easily invert for $\omega_{ab} = G_{ab}/(2+\bar \Pi^a_{a}+\bar \Pi^b_b)$. The kinetic terms for $B_i$ come entirely from $G_{0i} = \partial_t B_i - \dots$ and so to keep track of the sign if the kinetic term, we need only keep track of combinations that are quadratic in the pairs $G_{0i}$ and $\omega_{0i}$.

Focusing on only these terms we have (temporarily suspending Einstein summation convention)
\be
S_{\rm helicity-1/0}  = 2 \left( \sum_{n=0}^2 \frac{\hat \beta_{n+1}}{(2-n)! n!} \left( \frac{b}{a}\right)^n\right) \sum_i \left( G_{0i} \omega_{0i} - \frac{1}{2} \omega_{0i}^2 (2+ \bar \Pi^i_i + \bar \Pi^0_0) \right) + \dots \,
\ee
where we have used $1+ \bar \Pi^i_i = b/a$. Integrating out $\omega_{0i}$ gives
\be
S_{\rm helicity-1/0}  = \left( \sum_{n=0}^2 \frac{\hat \beta_{n+1}}{(2-n)! n!} \left( \frac{b}{a}\right)^n\right) \sum_i  \frac{(G_{0i})^2}{(2+ \bar \Pi^i_i + \bar \Pi^0_0)} + \dots
\ee
Since we always require the eigenvalues of $1+ \Pi$ to be positive definite for invertibility of the coordinate transformation between the two metrics we always have $(2+ \bar \Pi^i_i + \bar \Pi^0_0) > 0$. Thus the condition for the absence of ghosts in the vector sector is simply
\be
 \left( \sum_{n=0}^2 \frac{\hat \beta_{n+1}}{(2-n)! n!} \left( \frac{b}{a}\right)^n\right) > 0
\ee
which is equivalent to the statement that
\be
\tilde m^2(H) >0 \, .
\ee
This is a weaker condition that the generalized Higuchi bound, and is thus always satisfied whenever the generalized Higuchi bound is satisfied. For instance if all the $\hat \beta_n$ are positive it is guaranteed to be satisfied. We stress that this condition is not specific to de Sitter but works for any FRW geometries. It is precisely this condition that allows us to discard the branch of solutions given in Eq.~(\ref{branch2}).

\section{Partially Massless (Bi)Gravity}
   
\label{PM}

In de Sitter spacetime, the spin-2 unitary representations of the de Sitter group split into 3 categories, the massless graviton $m^2=0$ with 2 d.o.f., the massive gravitons with $m^2 > 2 H^2 $ with 5 d.o.f., and an additional representation $m^2=2H^2$ which carries only 4 degrees of freedom. This is known as the partially massless graviton \cite{Deser:1983mm,Deser:2001pe,Deser:2001us,Deser:2001xr,Deser:2004ji,Zinoviev:2001dt,Zinoviev:2006im,Deser:2006zx,Joung:2012hz,deRham:2012kf,Deser:2013xb,Deser:2013uy,Hassan:2012gz,deRham:2013wv}. As in the massless case, the field theory for the linear partially massless representation admits an additional symmetry which removes the 5th degree of freedom. The partially massless representation saturates the Higuchi bound in de Sitter. 

In \cite{deRham:2012kf} it was suggested that the dRGT model of massive gravity with a de Sitter reference metric could, for a special choice of the parameters in the potential ($\beta_n$) act as a candidate for a consistent nonlinear interacting theory of a single partially massless graviton. This conjecture was further extended to bigravity in \cite{Hassan:2012gz}. These are what we refer to as the candidate partially massless (bi)gravity theories.

Unsurprisingly, the bigravity and massive gravity cases are connected. The partially massless bigravity is defined by the choice  $\beta_0 = \frac{3}{2} \mpl^4/\mpf^2$, $\beta_2 = \mpl^2$, $\beta_4 = \frac{3}{2}  \mpf^2$, $\beta_1=\beta_3=0$.
Explicitly, the action for partially massless bigravity is given by \cite{Hassan:2012gz} 
\be
\label{version1}
\mathcal{L}=\frac{1}{2}\left[\mpl^2 \, \sqrt{-g} R[g] + \mpf^2\sqrt{-f}  R[f]-{m^2} \sqrt{-g} \( 3\frac{\mpl^4}{\mpf^2} \, {\cal U}_0 + \mpl^2 \, {\cal U}_2(X) + 3\mpf^2 \, {\cal U}_4(X)  \)  \right] \, ,
\ee
where $X = \sqrt{g^{-1}f}$. In this representation there are two different Planck masses. We may also choose to rescale one of the metrics as $g \rightarrow \mpf \tilde g/\mpl$ so that
\be
\label{version2}
\mathcal{L}=\frac{\mpf^2}{2}\left[  \sqrt{-\tilde g} R[\tilde g] +\sqrt{-f}  R[f]-{m^2 } \sqrt{-g} ( 3 \, {\cal U}_0(\tilde X) +  {\cal U}_2(\tilde X) +3  \, {\cal U}_4(\tilde X)  )  \right] \, ,
\ee
where $\tilde X = \sqrt{{\tilde g}^{-1}f}$. To emphasize the duality between the metrics, we may use the properties of the symmetric polynomials to write this as
\be
\label{version3}
\mathcal{L}=\frac{\mpf^2}{2}\left[  \sqrt{-\tilde g} R[\tilde g] +\sqrt{-f}  R[f]-{m^2 } \sqrt{-f} ( 3 \, {\cal U}_0(\hat X) +  {\cal U}_2(\hat X) +3  \, {\cal U}_4(\hat X)  )  \right] \, ,
\ee
where $\hat X = \sqrt{{ f}^{-1} \tilde g}$.
Although (\ref{version2}) and (\ref{version3}) are simpler, we choose to work in the asymmetric representation (\ref{version1}) with two different Planck masses so that we may more easily see the $\mpf/\mpl \rightarrow \infty$ decoupling limit to massive gravity. 

For partially massless bigravity, one combination of the scale factors becomes pure gauge. This is the extra symmetry that must arise in the partially massless case. 
The Friedmann equation for the $f$ metric is given by
\be
H_f^2 \mpf^2 =\frac{1}{2} m^2 \mpl^2 \(\frac{H_f^2}{H^2} + \frac{\mpf^2}{\mpl^2} \) \, . 
\ee
As already mentioned in the introduction, given this equation, the generalized Higuchi bound is saturated. This is remarkable since it is true for any FRW (i.e. it is independent of $\dot H$), and for any matter content coupled to the $g$ metric. The saturation of the bound is the indication that the kinetic term for the helicity-0 mode vanishes which is a necessary requirement for the existence of the partially massless theory. 

By taking the limit $\mpf \rightarrow \infty$ for fixed $\mpl$ and $H$ we see that this equation becomes simply
\be
H_f^2 = m^2/2 \, ,
\ee
which is the usual Higuchi condition imposed on the reference metric. Thus in the $\mpf/\mpl \rightarrow \infty$ decoupling limit, the proposed partially massless bigravity of \cite{Hassan:2012gz} becomes equivalent to the proposed partially massless gravity of \cite{deRham:2012kf}. This decoupling limit is well defined at the level of the action and the equations of motion. To see this at the level of the action, we begin with the partially massless bigravity action in the asymmetric representation 
\be
\mathcal{L}=\frac{1}{2}\left[\mpl^2 \, \sqrt{-g} R[g] + \mpf^2\sqrt{-f}  R[f]-{m^2} \sqrt{-g} \( 3\frac{\mpl^4}{\mpf^2} \, {\cal U}_0 + \mpl^2 \, {\cal U}_2(X) + 3\mpf^2 \, {\cal U}_4(X)  \)  \right] \, . \nn
\ee
We have seen from the Friedmann equation that we can consistently take the limit $\mpf \rightarrow \infty$ so that the $f$ metric asymptotes to de Sitter with Hubble constant $H_f^2 = m^2/2$. We generalize this by expressing $f_{\mu\nu} = f^{dS}_{\mu\nu} + v_{\mu\nu}/\mpf $ where $f^{dS}_{\mu\nu} $ is de Sitter with Hubble constant $H_f^2=m^2/2$. $v_{\mu\nu}$ denotes a perturbation around the de Sitter metric. Any metric $f$ may be expressed in a locally inertial de Sitter frame so that within a region of size $1/\sqrt{\delta R(f)}$ the metric takes this form.
Plugging this expression into the action we can take the limit $\mpf \rightarrow \infty$ after subtraction of a non-dynamical total derivative term. The resulting theory is
\bea
&& \lim_{\mpf \rightarrow \infty}\left( \mathcal{L} - \frac{1}{2}\mpf^2\sqrt{-f_{dS}}  R[f_{dS}]+ \frac{3}{2} m^2 \mpf^2 \sqrt{-f_{dS}} \right)  \nn \\ 
&=&\frac{1}{2}\left[\mpl^2 \, \sqrt{-g} R[g] -\frac{1}{2} v^{\mu\nu} (\hat {\cal E}_{dS} v)_{\mu\nu}  -{m^2} \sqrt{-g} \mpl^2 {\cal U}_2(X_{dS})    \right] \, ,
\eea
where $X_{dS} = \sqrt{g^{-1}f_{dS}}$ and $\hat {\cal E}_{dS}$ is the Lichnerowicz operator on the de Sitter reference metric $f_{dS}$, $\hat {\cal E}_{dS} = -\frac{1}{2}\Box_{f_{dS}} + \dots$. Thus we see that the $\mpf/\mpl \rightarrow \infty$ decoupling limit of partially massless bigravity is precisely equivalent to partially massless gravity and a decoupled massless graviton living on the reference de Sitter metric. 
The same limit may be taken at the level of the equations of motion where we find the same result. The latter of course follows from the fact that the limit is well defined at the level of the action. 

Unfortunately both this result and the result of {\it Section}~\ref{Dec} rule out the conjectured partially massless bigravity for the same reasons as given in \cite{deRham:2013wv} (see also \cite{Deser:2013uy} and \cite{Deser:2013gpa}). Even keeping the ratio of the two Planck scales fixed $\mpf/\mpl$, the $\Lambda_3$ decoupling limit of partially massless bigravity gives the same helicity-1/helicity-0 interactions as partially massless gravity. It is consistent to take this limit because even though $m \rightarrow 0 $, the Higuchi bound is not violated in the limit since we also scale $H^2 \sim R \sim 1/\mpl \sim m^2$.
The helicity-1/helicity-0 interactions are not affected by the dynamics of the second metric since the additional degrees of freedom represented by $v_{\mu\nu}$ are weakly coupled in this limit. 
The problem interactions in the candidate partially massless theories are the same in bigravity and massive gravity
\bea
S_{\rm helicity-1/0}  &=& - \frac{1}{8} \de^{\mu\nu\rho\si}_{abcd}  \left(  2 G_{\mu}^a (\de + \Pi)^b_{\nu} \om^c {}_{\rho}  \de^d_{\si}+  (\de + \Pi)^a_{\mu} (\de + \Pi)^b_{\nu} [  \om^c{}_{\rho}\om^d{}_{\si}+ \de^d_{\si}\om^c{}_{\al} \om^{\al}{}_{\rho}]   \right) \, \nn \\
\eea
where
\bea
\om_{ab}  &=&\int^{\infty}_{0} \d u \hspace{.2cm} e^{-2u} e^{ -u\Pi_a{}^{a'} } G_{a'b'} e^{ -u\Pi^{b'}_{\hspace{.2cm}b}}  \, ,
\eea
is the solution of 
\be
G_{ab} = \partial_a B_b - \partial_b B_a =\om_{ac}(\delta + \Pi)^c{}_b + (\delta + \Pi)_a{}^c  \om_{cb}\,.
\ee
There are no additional free parameters to adjust the cofficients of the interactions. The conjectured symmetry of partially massless (bi)gravity requires that the helicity-0 mode becomes pure gauge so that there are only 4 (in massive gravity) or 4+2 (in bigravity) dynamical degrees of freedom. However as argued in  \cite{deRham:2013wv} the presence of helicity-1/helicity-0 interactions negates this. In a nonzero background field for $G_{ab}$ a kinetic term will be generated for $\pi$ and there will be 5 or 5+2 propagating degrees of freedom. This implies that there is no hidden symmetry (it is essentially an accident of the linearized theory around backgrounds which contain no vector perturbations). It is plausible that there exists a modification to the kinetic term of the conjectured partially massless theories that resolves this issue.

\section{Conclusions}

In this work we set out to provide a simple and consistent derivation of the stability condition on the kinetic term of the helicity-0 mode in ghost-free massive gravity and bigravity theories, when both metrics are spatially flat FRW.
First, we confirmed the realization \cite{Fasiello:2012rw} that exact spatially flat FRW solutions in massive gravity with an FRW reference metric are ruled out on the basis of combined and, as it turns out, conflicting inequalities stemming from stability and observational constraints. This is what we call the \textit{Higuchi vs Vainshtein} (H-V) tension characterizing a natural set of candidate cosmological solutions for massive gravity.

We want to stress again that, although these solutions are appealing due to their simplicity, they are not the only way to achieve approximately FRW solutions in massive gravity \cite{massivec}. As such, this result does not in any way represent the last word on cosmological solutions in massive gravity. What we have ruled out is one specific idea of how to achieve a realistic cosmology in massive gravity. 

One of the possible escape routes from the H-V tension, one that does not require us to part from spatially flat FRW solutions, is represented by a closely related extension of massive gravity, ghost-free bigravity. We considered the generalization of the Higuchi bound to the known spatially flat FRW solutions in bigravity.  Interestingly, we have found that in bigravity theories it is indeed possible to relax the stability bound is a way that does not conflict with observational constraints because of the way the bound is modified: 
\bea
{\tilde m}^2(H) \left[ H^2+\frac{H_f^2 \mpl^2}{\mpf^2}\right] \geq 2H^4.
\eea
One can now choose to live in the $H_f/H\gg \mpf/\mpl$ region of the parameter space and this condition alone will satisfy the stability bound by virtue of the Friedmann equation for $H_f$.  No further constraint on the dressed mass $\tilde{m}$ or the bare mass $m$ is implied. 
 
We have provided two independent non-trivial checks of our result, obtaining the stability bound by building on a detailed analysis of the properties of the minisuperspace action and, later, by considering the powerful $\Lambda_3$ decoupling limit of massive gravity and bigravity which, in the bigravity case, we derive here for the first time explicitly. Using the decoupling limit, we find that the vector modes are automatically stable whenever the above bound is satisfied. 

Our derivation of the decoupling limit for bigravity has also led us to uncover a dual formulation of Galileons. In retrospect, we can easily say why this duality exists. It follows from the $f\leftrightarrow g$ symmetry of bigravity. In dRGT massive gravity Galilean interactions (the ubiquitous $\pi$ field) are identified as the helicity-0 mode in the map relating the  $g$ and $f$ metric coordinate system. Associated with this map is an inverse map which can be used to define the dual Galileon. The dual Galileon describes the way Galileons are seen from the perspective of the $f$ metric. We have shown here the existence of a non-local field redefinition that relates the Galileon $\pi$ to its dual Galileon field $\rho$. We discuss the duality in more detail in \cite{duality}. 

We provide yet another check on our result on the generalized stability bound by means of the partially massless bigravity theory. The latter is special in that, owing to a specific choice on the value of the $\beta_n$ coefficients (the same as in PM gravity) the stability bound is always saturated, irrespective of matter content, and the two Friedmann equations are degenerate implying that one of the scale factors becomes pure gauge.  We verified that, upon using the $f$ metric Friedmann equation in the PM limit of the bigravity stability bound, the inequality turns into an identity which is independent from both $H,H_f$. This then represents another non-trivial check on the generalized stability condition we unveiled. Unfortunately, we find that using the bigravity decoupling limit, the candidate partially massless bigravity models suffer the same problems in the vector sector as the candidate partially massless gravity, a fact which seems to preclude their existence, at least in their presently proposed form. 

In summary, the study of stable and viable cosmological spatially flat FRW solutions for ghost-free models of massive (bi)gravity has lead us to a generalized stability bound for these theories and has further pointed us towards uncovering and studying a dual formulation of Galileons which certainly deserve further investigation \cite{duality}.

\acknowledgments

We would like to thank Claudia de Rham, Rachel Rosen and Kurt Hinterbichler for discussions that suggested the minisuperspace approach to the bound. We are very grateful to  Andrew Matas and Nick A. Ondo for insightful comments on an early version of this manuscript.
AJT is supported by Department of Energy Early Career Award DE-SC0010600. AJT would also like to thank the Yukawa Institute for Theoretical Physics at Kyoto University,  ``Nonlinear Massive Gravity Theory and its Observational Test" workshop, YITP-T-12-04, where the results on the stability bound in bigravity were first presented \cite{presentation}. MF is supported in part by D.O.E. grant DE-SC0010600.

\section*{Appendix: Fluctuations in de Sitter Bigravity}
\label{Appendix}

In this article we have derived the analogue of the Higuchi bound in for FRW geometries for both massive gravity and bigravity. The most immediate way to our result, is to specialize to the case of linear fluctuations around de Sitter for both metrics. This argument is not as general as our previous results, and it gives little insight into the physics of the Vainshtein mechanism. Nevertheless, we present the analysis here to give an independent check on our calculations. 

Beginning with the vierbein form of the bigravity Lagrangian
\bea
S_{\rm bigravity} &=& \frac{\mpl^2}{2} \ep_{abcd} \int  \frac{1}{2}E^a \wedge E^b \wedge R^{cd}[E]+ \frac{\mpf^2}{2} \ep_{abcd} \int  \frac{1}{2}F^a \wedge F^b \wedge R^{cd}[f]  \nn  \\
&-& \frac{m^2}{2} \ep_{abcd} \int  \left[ \frac{\b_0}{4!} E^a \wedge E^b \wedge E^c \wedge E^d  + \frac{\b_1}{3!} F^a \wedge E^b \wedge E^c \wedge E^d  \right] \nonumber \\
&-& \frac{m^2}{2} \ep_{abcd} \int \left[\frac{\b_2}{2!2!} F^a \wedge F^b \wedge E^c \wedge E^d  + \frac{\b_3}{3!} F^a \wedge F^b \wedge F^c \wedge E^d \right]  \, .
\eea
we perturb each vierbein around the vierbein for de Sitter $V^a$ with Hubble constant $H$ so that the de Sitter metric is $\d s^2 = V_{\mu}^a V_{\nu}^b \eta_{ab} \d x^{\mu}\d x^{\nu} $ according to
\bea
&& E^a_{\mu} = V^a_{\nu} \( \delta^{\nu}_{\mu} + \frac{1}{2 \mpl} h^{\nu}_{\mu} \) \\
&& F^a_{\mu} = \frac{H}{H_f}V^a_{\nu}  \( \delta^{\nu}_{\mu} + \frac{1}{2 \mpf} \frac{H_f}{H} w^{\nu}_{\mu} \)
\eea
Then to second order in perturbations we have
\bea
S &=& \int \d^4 x \sqrt{-V} \left(-\frac{1}{4} h^{\mu\nu} \hat {\cal E}_{dS} h_{\mu\nu} -\frac{1}{4} w^{\mu\nu} \hat {\cal E}_{dS} w_{\mu\nu} \right.  \nn  \\
&& \left.  -\frac{1}{8}F_1 (h_{\mu \nu}h^{\mu\nu}-h^2) -\frac{1}{4}F_2 (h_{\mu \nu}w^{\mu\nu}-hw) -\frac{1}{8}F_3 (w_{\mu \nu}w^{\mu\nu}-w^2) \right)
\eea
where $\hat {\cal E}_{dS} $ is the Lichnerowiz operator on the de Sitter metric with Hubble $H$ and we raise and lower indices with respect to the same metric. 

The coefficients are given by
\bea 
F_1 &=&  3 m^2 H^2- \frac{m^2}{\mpl^2} \sum_n \frac{\beta_n}{(2-n)! n!} \left( \frac{H}{H_f}\right)^n  \, , \\
F_2 &=& -\frac{m^2}{\mpl \mpf} \sum_n \frac{\beta_n}{(3-n)! (n-1)!} \left( \frac{H}{H_f}\right)^{n-1} \, , \\
F_3 &=& 3 m^2 H^2 -\frac{m^2}{\mpf^2} \sum_n \frac{\beta_n}{(4-n)! (n-2)!} \left( \frac{H}{H_f}\right)^{n-2}  \, .
\eea
The term $3 m^2 H^2$ arises from the Einstein-Hilbert term, and is the one that is usually cancelled against the cosmological constant in GR. 
In the present case $\beta_0$ and $\beta_4$ (here nonzero) are fixed by the tadpole cancellation requirement, i.e. the requirement that the two de Sitter spacetimes are solutions. 
These can be inferred from the Friedmann equations
\bea
&&H^2 = \sum_{n=0}^3 \frac{\beta_n}{\mpl^2(3-n)! n!} \left( \frac{H}{H_f} \right)^n \, , \\
&&H_f^2 = \sum_{n=0}^3 \frac{\beta_{n+1}}{\mpf^2(3-n)! n!} \left( \frac{H}{H_f} \right)^{n-3} \, . 
\eea
Using these for $\beta_0$ and $\beta_4$ values and substituting back in we obtain
\bea
F_1 &=&  \tilde{m}^2(H) \, , \\
F_2 &=& -\frac{\mpl}{\mpf} \frac{H_f}{H} \tilde{m}^2(H)  \, ,\\
F_3 &=&  \frac{\mpl^2}{\mpf^2} \frac{H_f^2}{H^2} \tilde{m}^2(H) \, .
\eea
where we have used the same definition of dressed mass as in the main text.
These satisfy the zero mass eigenvalue condition $F_1 F_3-F_2^2=0$. The non-zero mass eigenvalue is given by
\be
m^2_{\rm eigen} = F_1+F_3 = \tilde m^2(H) \left( 1+\frac{\mpl^2}{\mpf^2} \frac{H_f^2}{H^2}  \right) \, .
\ee
Demanding the normal Higuchi conditon $m^2_{\rm eigen} \ge 2H^2$ gives the result derived in the text
\be
m^2_{\rm eigen} = \tilde m^2(H) \left( 1+\frac{\mpl^2}{\mpf^2} \frac{H_f^2}{H^2}  \right) \ge 2 H^2 \, .
\ee
Let us stress again though that the result in the main text is far more general, since it is valid for FRW geometries and arbitrary matter content. 

For the self-accelerating stable branch described in the {\it Section} \ref{cosmology} we have
\be
m^2_{\rm eigen} \sim 3 H^2 \, ,
\ee
for all times. Thus the physical mass of the massive graviton mode always remains close to the Hubble scale at a given time, in such a way as to ensure the bound is satisfied. 

%%%%%%%%%%%%%%%%%%%%%%%%%%%%%%%%%%%%%%%%%%%%%%%%%%%%%%%%%%%%%%%%%%%%%
%%%% Bibliography

%\newpage

%Unused bibitems

%\bibitem{Deffayet:2009mn} 
%  C.~Deffayet, S.~Deser and G.~Esposito-Farese,
%  ``Generalized Galileons: All scalar models whose curved background extensions maintain second-order field equations and stress-tensors,''
%  Phys.\ Rev.\ D {\bf 80}, 064015 (2009)
%  [arXiv:0906.1967 [gr-qc]].
%
%\bibitem{Hordenski}
%G. W. Horndeski, Int. J. Theor. Phys. 10 (1974).
%\bibitem{Deffayet:2010zh} 
%  C.~Deffayet, S.~Deser and G.~Esposito-Farese,
%  ``Arbitrary $p$-form Galileons,''
%  Phys.\ Rev.\ D {\bf 82}, 061501 (2010)
%  [arXiv:1007.5278 [gr-qc]].
%
%Unused bibitems

%
\end{document}